\documentclass[journal,onecolumn,11pt,times]{IEEEtran}
\pdfoutput=1 
\usepackage{amsmath}
\usepackage{fullpage}
\usepackage{graphicx}
\usepackage{amsmath}
\usepackage{amsfonts}
\usepackage{amssymb}
\usepackage{eucal}
\usepackage{subfigure}
\usepackage{multirow}
\usepackage{dblfloatfix}
\usepackage{color}
\usepackage{setspace}
\usepackage{citesort}
\usepackage[numbers,sort&compress]{natbib}

\graphicspath{{figs//}}
\title{On the Impact of Phase Noise on Active Cancellation in Wireless
  Full-Duplex}
\author{Achaleshwar Sahai, Gaurav Patel, Chris Dick and Ashutosh
  Sabharwal\footnote{Chris Dick is with Xilinx, Inc. Achaleshwar Sahai, Gaurav Patel
    and Ashutosh Sabharwal are in the Department of Electrical and Computer
    Engineering at Rice University. Their work was partially supported by an NSF
    EAGER Grant CCF 1144041 and a grant from Xilinx, Inc. The results in
    this paper was presented in part at the Asilomar Conference on
    Signals, Systems, and Computers, 2012 \cite{asilomar2012}.}}

\date{}

\begin{document}

\maketitle
\vspace{-0.5in}
\begin{abstract}
Recent experimental results have shown that full-duplex communication
is possible for short-range communications. However, extending
full-duplex to long-range communication remains a challenge, primarily
due to residual self-interference even with a combination of passive
suppression and active cancellation methods. In this paper, we
investigate the root cause of performance bottlenecks in current
full-duplex systems. We first classify all known full-duplex
architectures based on how they compute their cancelling signal and
where the cancelling signal is injected to cancel
self-interference. Based on the classification, we analytically
explain several published experimental results. The key bottleneck in
current systems turns out to be the phase noise in the local
oscillators in the transmit and receive chain of the full-duplex
node. As a key by-product of our analysis, we propose signal models
for wideband and MIMO full-duplex systems, capturing all the salient
design parameters, and thus allowing future analytical development of
advanced coding and signal design for full-duplex systems.
%
%We show that the variance of phase noise limits the amount of
%active analog cancellation. We also show that the residual after
%active analog cancellation scales with the variance of phase noise as
%well as the strength of self-interference channel. A serially
%concatenated digital canceller does not cancel the residual dependent
%on phase noise. Thus, the total amount of cancellation of active
%analog and digital stages is limited by the variance of phase
%noise. We also show that in pre-radio cancellers, due to phase noise
%the amount of active cancellation is influenced by the amount of
%passive suppression. We show that although increasing the amount of
%passive suppression improves the sum total of suppression and
%cancellation, the amount of active cancellation reduces.
\end{abstract}

\section{Introduction}
\label{sec:intro}
In full-duplex communication, a node can simultaneously transmit one
signal and receive another signal on the same frequency band. The key
challenge in full-duplex communications is the
\emph{self-interference}, which is the transmitted signal being added
to the receive path of the same node. Due to the proximity of the
transmit and receive antennas on a node, self-interference is often
many orders of magnitude larger than the signal of interest. Thus, the
main objective for full-duplex design is to reduce the strength of
self-interference as much as possible -- ideally, down to noise floor.

%Proximity of
%the transmit and receive antenna on the full-duplex node results in
%interference, called \emph{self-interference}, which is several orders
%of magnitude higher than the signal being received. The main challenge
%in enabling full-duplex is managing self-interference in such a way
%that allows reliable decoding of the signal being received.

Self-interference is usually reduced by a combination of passive and
active
methods~\cite{FDmicrosoft,Choi:2010aa,khandani,Duarte:2010aa,Choi:2011m,achal:fd,Khojastepour:2011,midu,Duarte:phd,duarte2012}. Passive
methods, which use antenna designs, aim to \emph{increase} the
pathloss for the self-interference signal. In contrast, active methods
employ the knowledge of self-interference to cancel it from the
received signal. However, none of the designs
\cite{Choi:2010aa,Duarte:2010aa,Choi:2011m,achal:fd,Khojastepour:2011,midu,khandani,FDmicrosoft,Duarte:phd,duarte2012}
manage to eliminate self-interference completely. In fact,
in~\cite{Duarte:phd}, authors report that even after passive
suppression and active cancellation, the strength of self-interference
is 15~dB above the thermal noise floor. Our main focus, in this paper,
is to understand the bottlenecks that limit self-interference from
being completely eliminated in current full-duplex systems by
answering the following three questions, observed experimentally in
prior works.

\emph{Question 1}: Active cancellation can occur before or after
analog-to-digital conversion. If active cancellation occurs prior to
digitization of the received signal, it is referred to as active
analog cancellation. The cancellation that operates on the received
signal in digital baseband is labeled  digital cancellation. Designs
\cite{Choi:2010aa,Duarte:2010aa,Choi:2011m,Khojastepour:2011} report
anywhere between 20-45~dB of active analog cancellation, which raises
the first question that we answer analytically in this paper ``What
limits the amount of active analog cancellation in a full-duplex system
design?''

\emph{Question 2}: An interesting observation reported in
\cite{Duarte:phd} is that if active analog cancellation and digital
cancellation are cascaded together, then the amount of digital
cancellation depends on the amount of analog cancellation. More
specifically, \cite{Duarte:phd} reports that whenever their analog
canceller cancels less self-interference, then the digital canceller
cancels more and vice versa. The above observation leads to the second
question which we answer, ``How do the amounts of cancellations by
active analog and digital cancellers depend on each other in a
cascaded system?''

\emph{Question 3}: 
Finally, in \cite{Duarte:phd}, it is also reported that more passive
suppression results in increased total self-interference reduction, when  both passive suppression and active analog cancellation are used. However, the total reduction does not increase linearly with amount of passive suppression.
%In contrast, some of the past work~\cite{Choi:2010aa,Choi:2011m,Khojastepour:2011} assumed that passive suppression is independent of active analog cancellation. 
We make preliminary progress towards answering the third question,
``How and when does passive suppression impact the amount of active
analog cancellation?''

%% {\bf Achal, I think for this paper, we should confuse the reader
%% with three contributions which do not match the three
%% questions. Instead, one approach will be that say 'In this paper,
%% we answer all three answers using the following procedure. First,
%% we harmonize all known architectures. Then we show that phase noise
%% answers all questions. To answer Question 1, .. To answer Question
%% 2.... To answer Question 3... Finally, as a side product of our
%% work, ...}

In this paper, we answer all the three questions using the following
procedure. First, we harmonize all known architectures of active
analog cancellers by classifying them into two classes: pre-mixer and
post-mixer cancellers, based on where the cancelling signal is
generated. Pre-mixer canceller \cite{Duarte:2010aa} generates the
cancelling signal prior to upconversion in the digital baseband, while
post-mixer canceller \cite{Choi:2010aa,Choi:2011m,Khojastepour:2011}
generates the self-interference signal at the carrier frequency. Both
pre- and post-mixer perform the cancellation at the carrier
frequency. As a side result, our classification yields another active
analog canceller architecture which we label as baseband analog
canceller. In baseband analog cancellation, both the cancelling signal
as well as cancellation operation is in the \emph{analog}
baseband. The above classification of analog cancellers allows us to
study all architectures systematically using one umbrella analysis,
and thus allows direct comparisons between performance of different
cancellers.

Once we classify the known architectures of full-duplex designs, we
show that phase noise~\cite{ad:phasenoise} associated with local
oscillators at the transmitter and receiver turns out to be the source
of major bottleneck in full-duplex systems. In fact, phase noise
answers all three questions raised above. To answer Question~1, we
analyze the amount of active analog cancellation possible in different
types of cancellers and show that by incorporating phase noise into
the signal model, we can closely match the cancellation number
reported in \cite{Duarte:2010aa} and conjecture that phase noise also
explains results of~\cite{Choi:2011m,Khojastepour:2011}.

To answer Question~2, we show that the amount of active analog
cancellation and concatenated digital cancellation is limited by a
quantity that depends on the phase noise properties of the local
oscillators. We show that, if the active analog canceller cancels
more, the residual self-interference has a dominant contribution of
phase noise, which is uncorrelated to the self-interference signal and
thus cannot be cancelled by the concatenated digital canceller. On the
other hand, if active analog canceller cancels less, the residual
self-interference has a higher correlation to the self-interference
signal and thus a larger fraction of self-interference can be
cancelled by the digital canceller.

To answer Question~3, we show that due to phase noise the amount of
active analog cancellation, in a pre-mixer canceller, is dependent on
the amount of passive suppression. We show that the sum total of
passive suppression and active analog cancellation increases with an
increase in passive suppression, but individually the amount of active
analog cancellation reduces as the amount of passive suppression
increases.  As a result, the sum total of passive suppression and
active cancellation does not increase linearly with increase in
passive suppression.

Finally, as a by-product of our analysis of active cancellers, we
propose signal models for MIMO and wideband full-duplex systems. The
signal models allow us to abstract away the form of active
cancellation, and can be used for signal design and analysis of
full-duplex systems. The noise term in the proposed signal model
depends on three parameters: phase noise variance and its
autocorrelation, quality of self-interference channel estimates and
thermal noise. Each of the three parameters decides the dominant noise
in full-duplex system in different regimes of transmitted
self-interference power, thus captures the limits of communication in
full-duplex.

The rest of the paper is organized as follows. In
Section~\ref{sec:Cancelation}, we review the need for active
cancellation and classify the different known architectures of active
analog cancellers. In Section~\ref{sec:FirstAttempt}, we show that
self-interference channel estimation error does not explain the amount
of active analog cancellation reported in literature
\cite{Duarte:2010aa,Choi:2010aa,Choi:2011m,Khojastepour:2011}. In
Section~\ref{sec:experiment}, via a controlled experiment, we show
that phase noise limits the amount of active cancellation. In
Section~\ref{sec:AllCancellers} and \ref{sec:DigitalCancellation}, we
analyse the amount of active analog cancellation and concatenated
digital cancellation possible in different cancellers, uncovering
their interdependence. In Section~\ref{sec:PassiveVsActive}, we show
the interdependence between passive suppression and active
cancellation for pre-mixer cancellers. Finally, in
Section~\ref{sec:Model} we propose the MIMO and wideband signal model
for full-duplex systems. We conclude in Section~\ref{sec:conclusions}.

\section{Reducing Self-Interference in Full-Duplex}
\label{sec:Cancelation}
\subsection{Need for self-interference reduction}
Due to simultaneous transmission and reception in full-duplex, a
combination of incoming signal of interest and self-interference is
received at the full-duplex node.  Since the transmit and receive
antenna at the full-duplex node are in physical proximity, the
self-interference signal can be 50-100 dB larger
in magnitude compared to the signal of interest. For baseband
processing, the received signal is digitized using an analog to
digital convertor, which has a finite number of bits of
quantization. Before digitizing the signal, the automatic gain
control scales the input to a nominal range of
$[-1,1]$. Since the signal of interest is weaker than the
self-interference, the gain control settings are largely governed by the
strength of the self-interference, leading to the signal of interest
occupying a range much smaller than $[-1,1]$ in the quantized signal.
%% {\bf Achal, the text after this point is all over the place. You first
%%   say that post-quantization cancelation, the SNR will be low. Then
%%   your last sentence seems to say that removing interference in
%%   digital domain will work.}
After digitization, even if the self-interference signal can be
perfectly subtracted out, the quantization noise for the signal of
interest will be significantly large, leading to a very low effective
SNR in digital baseband. Thus, it is important to reduce the
self-interference prior to analog to digital conversion, so that the signal of interest will have a better
effective SNR in digital baseband.
%%  If the self-interference is not
%% completely eliminated prior to digitization, then the remaining
%% self-interference can be potentially reduced via active methods in
%% digital baseband, thus enabling reliable decoding of the signal of
%% interest.

\subsection{Methods of reducing self-interference}
\label{sec:ReducingSI}
\begin{figure}[ht]
\begin{minipage}[b]{0.45\linewidth}
  \centering
  \scalebox{0.5}{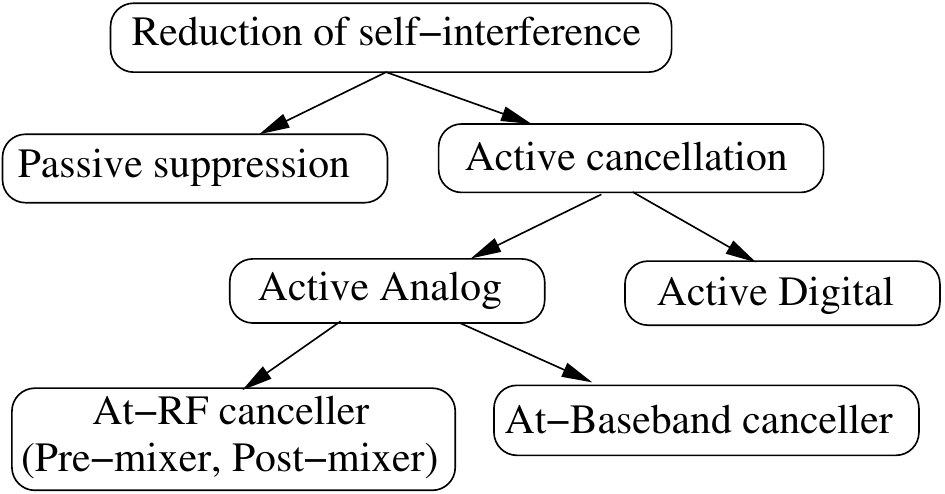}
  \caption{Classification of methods of reducing self-interference}
  \label{fig:chart}
\end{minipage}
\hspace{1cm}
\begin{minipage}[b]{0.45\linewidth}
\centering
  \scalebox{0.6}{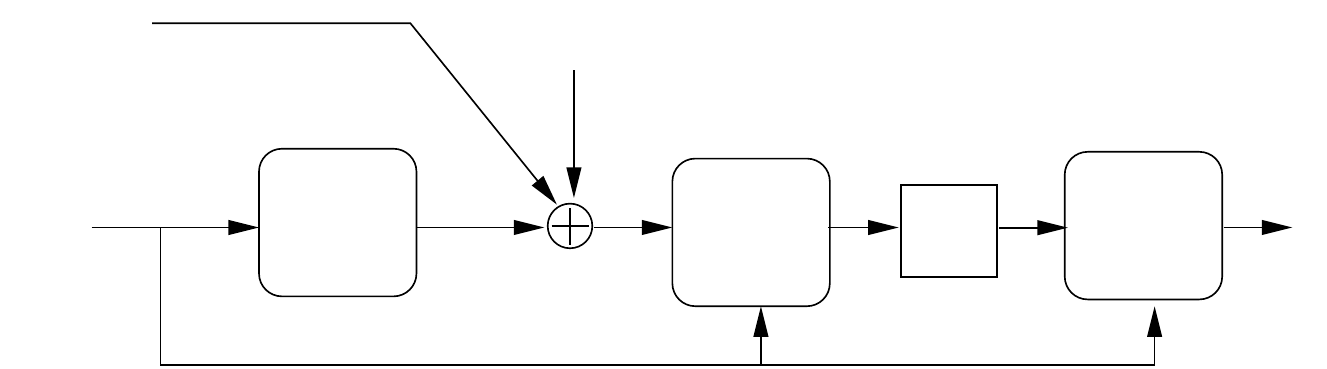}
  \caption{Block diagram representation of all the self-interference
    reduction methods in concatenation.}
    \label{fig:blkdia}
\end{minipage}
\end{figure}

Self-interference is reduced by both passive and active techniques.  A
diagramatic classification of methods of reducing self-interference is
shown in Figure~\ref{fig:chart}. A figure of merit to characterize any
technique used to reduce self-interference is the ratio of the
strength of self-interference before and after the technique is
employed, which is called the amount of suppression for passive and
cancellation for active techniques. Following is a brief review of the
methods to reduce self-interference.

\subsubsection{Passive suppression}
Passive suppression aims to reduce the self-interference by reducing
the electromagnetic coupling between the transmit and receive antenna
at the full-duplex node. As shown in Figure~\ref{fig:blkdia}, the
reduction in the strength of self-interference via passive methods
occurs before the self-interference signal impinges upon the receive
antenna. Passive methods include (a) \emph{antenna-separation}, which
achieves reduction of the self-interference by increasing pathloss
between transmit and receive antenna
\cite{achal:fd,Duarte:2010aa,Choi:2010aa}, (b)
\emph{directional-separation}, where the transmit and receive antenna
on the full-duplex node have lower mutual coupling as the main lobes
of the antennas do not point to each other \cite{evan,evanmaster}, (c)
\emph{polarization decoupling} \cite{evanmaster}, where the transmit
and receive antenna operate on orthogonal polarizations to reduce the
coupling.

\subsubsection{Active analog cancellation}
\label{sec:ActiveAnalogCanceller}
The mechanism of reducing self-interference which employs the
knowledge of self-interference to actively inject a cancelling signal
into the received signal in the analog domain is referred to as active
analog cancellation. As shown in Figure~\ref{fig:blkdia}, active
analog cancellation operates on the received signal. Thus, active
analog cancellation occurs after passive suppression. The objective of
active analog cancellation is to create a null for the
self-interference signal. The null for self-interference can be
created by performing cancellation either at the carrier frequency
(RF) or at the analog baseband. Most active analog cancellers
\cite{Choi:2010aa,Duarte:2010aa,Choi:2011m,Khojastepour:2011} cancel
self-interference at RF. We first classify active analog cancellers which cancel at RF and then describe the canceller
which cancel in analog baseband.

\begin{figure}[htb]
\label{fig:AnalogCanceller}
\centering \subfigure[At-RF active analog canceller]{\label{fig:AnalogCancellerRF}\scalebox{0.5}{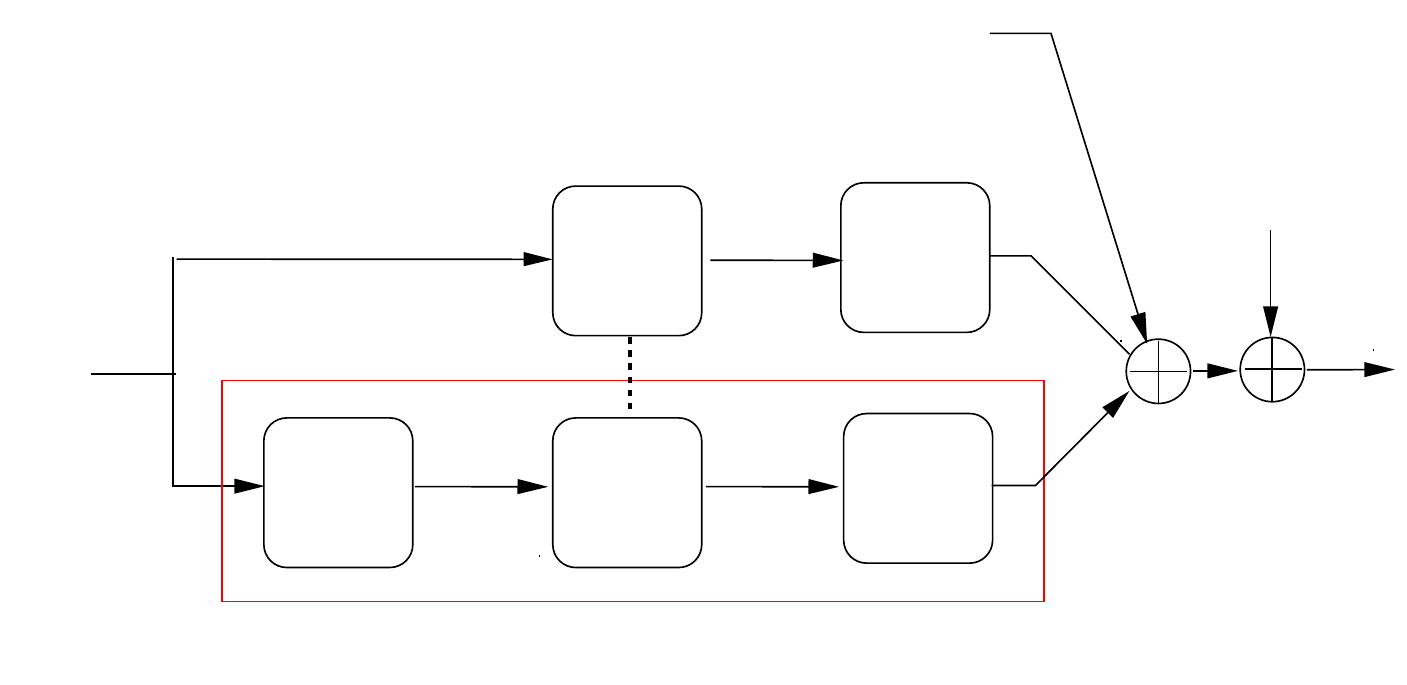}
} \hspace{10mm} \subfigure[At-Baseband active analog canceller.]{\label{fig:AnalogCancellerNoRF}\scalebox{0.5}{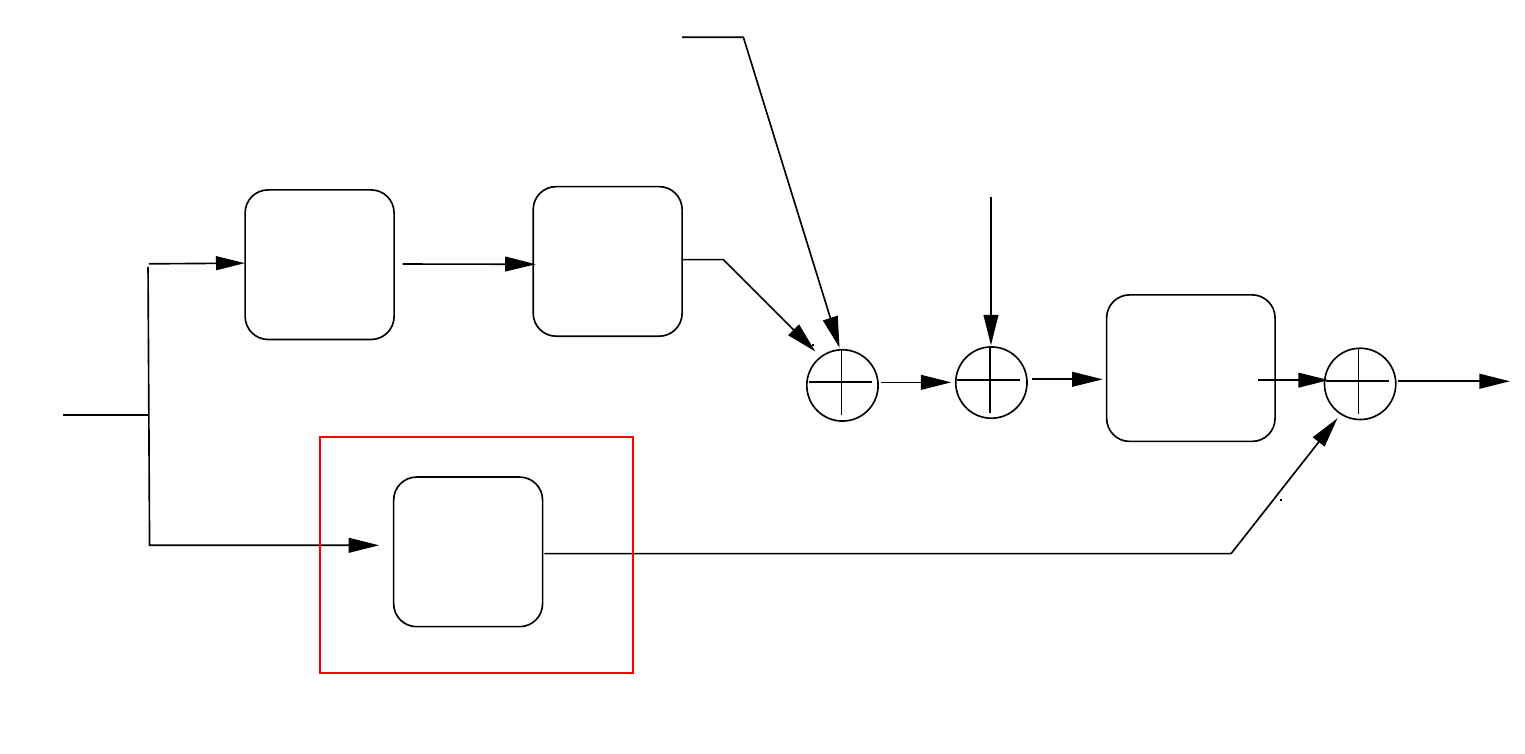}}
\caption{Two architectures of analog cancellers differentiated based
  on whether the cancellation occurs at RF or analog baseband. The
  functions $r_{\sf up}(.)$ and $r_{\sf down}(.)$ represent the
  process of upconversion to RF and downconversion from RF respectively.}
\end{figure}

\paragraph{Active analog cancellation at RF} In Figure~\ref{fig:AnalogCancellerRF},
we depict a block diagram of an active analog canceller which cancels
at RF. Note that, if the cancellation has to be performed at RF, then
the cancelling signal also needs to be upconverted to RF. The
cancelling signal is generated by processing the self-interference
signal $x_{\sf si}(t)$. We classify active analog cancellers based on
whether the cancelling signal has been generated by processing the
self-interference signal $x_{\sf si}(t)$, prior or post upconversion.
Those cancellers where the cancelling signal is generated by
processing $x_{\sf si}(t)$ prior to upconversion are called pre-mixer
cancellers, while cancellers where the cancelling signal is generated
by processing after $x_{\sf si}(t)$ is upconverted are called
post-mixer cancellers. Figure~\ref{fig:AnalogCancellerRF} shows the
pre-mixer processing function $f(.)$ and post-mixer processing
function $g(.)$. The choice of functions $f(.)$ and $g(.)$ are ideal
if after cancellation the self-interference signal is completely
eliminated from the received signal. For many known implementations,
we show the choice of functions $f(.)$ and $g(.)$ classify them as
pre- and post-mixer cancellers as follows.

\noindent
 \textbf{Parallel radio cancellation}: In \cite{Duarte:2010aa}, the
 negative of the scaled self-interference signal being received at the
 receiver of the full-duplex node is generated in the digital baseband
 and unconverted via a parallel radio chain.  The cancelling signal is
 then added to the received signal at carrier frequency using a
 passive power combiner. The functions
\begin{equation}
f(t) = h(t); g(t) = \delta(t),
\end{equation}
are implemented as filters, where $h(t)$ is a filter that is
implemented in digital domain. To cancel the self-interference, the
design implements $h(t) = -\hat{h}_{\sf si}(t)$, where $\hat{h}_{\sf
  si}(t)$ is the estimate of the self-interference channel $h_{\sf
  si}(t)$. If $\hat{h}_{\sf si}(t) = h_{\sf si}(t)$, and $\ast$
represents the convolution operation, then the cancellation should
result $h_{\sf si}(t)\ast x_{\sf si}(t) - \hat{h}_{\sf si}(t) \ast
x_{\sf si}(t) = (h_{\sf si}(t) - \hat{h}_{\sf si}(t))\ast x_{\sf
  si}(t) = 0$.

%% \item \textbf{Post-radio canceller} $\#2$ In \cite{Choi:2010aa}, at
%%   the full-duplex node $\mathsf{N}1$, two antennas are used to
%%   transmit. Let $\lambda_c$ be the wavelength corresponding to the
%%   carrier frequency. The transmitting antennas are placed $d$ distance
%%   apart. Both antennas transmit identical signals.
%% \begin{equation}
%%   f(t) = \delta(t); g(t) =
%%   e^{j\frac{2\pi d}{\lambda}}\delta(t),
%% \end{equation}
%% where $\lambda$ is the actual wavelength of transmission, and
%% $\delta(.)$ denotes the kronecker delta function. If $d =
%% \lambda_{c}/2$, and $\lambda = \lambda_{c}/2$, then the two copies of
%% self-interference signals which 180$^\circ$ out of phase will combine
%% creating a perfect null at receiver of $\sf N$1

\noindent
 \textbf{BALUN cancellation}: In \cite{Choi:2011m}, a copy of the
  signal in RF is passed through a BALUN\footnote{BALUN is a balanced
    unbalanced transformer, a single input two output device which
    converts signal balanced about to signal that is unbalanced} which
  produces the negative of the analog signal being transmitted. The
  negative signal is then amplified and delayed using a QHX220 analog
  chip~\cite{qhx220}, and finally added to the received signal in the
  analog domain, thus cancelling the self-interference. The generation
  of cancelling signal as well as cancellation occurs at carrier
  frequency, thus we classify BALUN cancellation as post-mixer
  cancellation. The functions
\begin{equation}
f(t) = \delta(t); g(t) = -g_1\delta(t) - g_2\delta(t - \tau),
\end{equation}
where $g_1$ and $g_2$ are gain coefficients and $\tau$ is a fixed
delay. If the coefficients $g_1$ and $g_2$ are chosen such that
$g_1\delta(t) + g_2\delta(t - \tau) = h_{\sf si}(t)$, then a null is
created at the receiver.

\noindent
 \textbf{Antenna cancellation}: In \cite{Khojastepour:2011}, at
  the full-duplex node, two transmit antennas Tx1$_a$ and Tx1$_b$ are
  placed at equal distance symmetrically away from the receive
  antenna. The transmit antennas transmit signals which are negative
  of each other. Upon reception, the copies of self-interference
  signals negate each other resulting in a smaller
  self-interference. Antenna cancellation is an example of post-mixer
  canceller because the processing occurs at RF as described by the
  functions
\begin{equation}
f(t) = \delta(t); g(t) = -h_{b, {\sf si }}(t),
\end{equation}
where $h_{b, {\sf si}}(t)$ is the over the air channel from antenna
Tx1$_b$ to the receive antenna. If the channel from Tx1$_{a}$ to the
receiver, $h_{a, {\sf si}}(t) = h_{b, {\sf si }}(t)$, then a perfect
null is created at the receiver.

Note that, in all the mechanisms described above, while the
cancellation is performed in RF-analog domain, the input to $f(.)$ can
either be a digital or an analog signal, while the input to $g(.)$ is
necessarily an analog signal.

\paragraph{Baseband analog canceller}
An active analog canceller where the cancelling signal is generated in
baseband as well as the cancellation occurs in the analog baseband is
called baseband analog canceller. Figure~\ref{fig:AnalogCancellerNoRF}
shows a representation of baseband analog canceller. In baseband
analog cancellers the self-interference signal $x_{\mathsf{si}}(t)$ is
processed by a function $s(.)$, either in baseband analog domain or in
digital domain before it is added to the received signal to perform
the cancellation. If the function $s(.)$ is such that the
self-interference signal is negative of the cancelling signal at the
receiver, then a null is created for the self-interference. Since the
cancelling signal does not go through upconversion process, possibly
less RF hardware is required to implement a baseband analog canceller.

\subsubsection{Digital cancellation}
The active cancellation which occurs in the digital domain after the
received signal has been quantized by an analog to digital convertor
is called active digital cancellation. Examples of full-duplex
systems where digital cancellation has been implemented are
\cite{Duarte:2010aa,Choi:2010aa}. From Figure~\ref{fig:blkdia}, we see
that digital cancellation is the final step of reduction of
self-interference. %Apart from operating in the digital domain,
digital

\section{First Attempt}
\label{sec:FirstAttempt}
In this section, we show that the conventional signal model for
narrowband communication does not satisfactorily explain the amount of
active analog %% {\bf Achal, analog or digital ?}  
cancellation reported in
\cite{Choi:2010aa,Duarte:2010aa,Choi:2011m,Khojastepour:2011}.
\subsection{Narrowband Signal Model}
Let $\sf N$1 denote a full-duplex node which transmits the
self-interference signal $x_{\sf si}(t)$, while $\sf N$2 denote the
node from which $\sf N$1 is receiving the signal of interest denoted
by $x_{\sf signal}(t)$. The impulse response of the self-interference
channel is denoted by $\mathbf{h_{\sf si}}(t)$, while the impulse
response of channel from $\sf N$2's transmitter to $\sf N$1's receiver
be denoted by $\mathbf{h_{\sf signal}}(t)$. Then, the received signal
at $\sf N$1 denoted by $y_1(t)$ is given by
\begin{equation}
y_1(t) = \mathbf{h}_{\sf si}(t)\ast x_{\sf si}(t) + \mathbf{h}_{\sf signal}(t) \ast x_{\sf signal}(t) + z_{\sf noise}(t),
\label{eq:BasicModel}
\end{equation}
where $\ast$ denotes the convolution operation, $z_{\sf noise}(t)$ is
the AWGN thermal noise distributed as $\mathcal{N}(0, \sigma_{\sf
  noise}^2)$. For a narrowband signal, the wireless channel can be
modeled as single tap delay channel, $\mathbf{h}_{\sf si}(t) = h_{\sf
  si}\delta(t - \Delta_{\sf si})$, and $\mathbf{h}_{\sf signal}(t) =
h_{\sf signal}\delta(t - \Delta_{\sf signal})$. Note that $h_{\sf si}$
and $h_{\sf signal}$ are complex attenuations which depend on channel
conditions, while $\Delta_{\sf si}, \Delta_{\sf signal} \in
\mathbb{R}^+$ 
%% {\bf Achal, can the delays be zero. If yes, then does
%%   $\mathbb{R}^+$ include 0? response: I don't think we should have the zero delay 
%% case}
 are delays with which the self-interference signal and the signal of
 interest, respectively, arrive at the receiver. Note that, the signal
 model in \eqref{eq:BasicModel} describes a time-invariant system. The
 assumption of time-invariance is valid as long we assume that
 \eqref{eq:BasicModel} describes $y_1(t)$ within the coherence times
 of the channels $\mathbf{h}_{\sf si}(t)$ and $\mathbf{h}_{\sf
   signal}(t)$. The average power at each of the transmitters is
 nominally limited to 1, which implies
\begin{equation}
\mathbb{E}(|x_{\sf si}(t)|^2) \leq 1,\text{ } \mathbb{E}(|x_{\sf
  signal}(t)|^2) \leq 1.
\end{equation}
The digital baseband equivalent of \eqref{eq:BasicModel} can be
written by replacing $t$ by $iT$ where $T$ is the sampling period and $i
\in \mathbb{Z}$.
\subsection{Amount of cancellation}
Let $\widehat{\mathbf{h}}_{\sf si}(t) = \hat{h}_{\sf si}\delta(t -
\hat{\Delta}_{\sf si})$ be the estimate of the self-interference
channel. With imperfect estimate of the channel, the residual
self-interference after active analog cancellation will be
\begin{equation}
  y_{1, {\sf residual}}(t) = h_{\sf si}x(t - \Delta_{\sf si}) - \hat{h}_{\sf si}x(t - \hat{\Delta}_{\sf
    si})  + z_{\sf noise}(t).
\label{eq:Cancel1}
\end{equation}
Equation \eqref{eq:Cancel1} implies that when
$\widehat{\mathbf{h}}_{\sf si}(t) = \mathbf{h}_{\sf si}(t)$, then the
residual is only due to thermal noise. The strength of the residual
self-interference is given by
\begin{eqnarray}
  \sigma_{\sf residual}^2 & = & \mathbb{E}(|y_{1, {\sf
      residual}}(t)|^2) \nonumber \\
 & \stackrel{\text{(a)}}{=} &
  \mathbb{E}\left(|\hat{h}_{\sf si}x(t - \hat{\Delta}_{\sf si}) - h_{\sf
    si}x(t - \Delta_{\sf si})|^2\right) + \sigma_{\sf noise}^2 \nonumber \\
  & = & \mathbb{E}\left(|\hat{h}_{\sf si}(x_{\sf si}(t -
  \hat{\Delta}_{\sf si}) - x_{\sf si}(t - \Delta_{\sf si})) +
  (\hat{h}_{\sf si} - h_{\sf si})x_{\sf si}(t - \Delta_{\sf si})|^2\right) +
  \sigma_{\sf noise}^2 \nonumber \\
 & \stackrel{\text{(b)}}{=} &
  2\mathbb{E}\left(|\hat{h}_{\sf si}|^2\right)(1 - R_{x_{\sf
      si}}(\hat{\Delta}_{\sf si} - \Delta_{\sf si})) +
  \mathbb{E}\left(|\hat{h}_{\sf si} - h_{\sf si}|^2\right) 
\nonumber
\end{eqnarray}
\begin{eqnarray}
&& + 2\text{Re}\left\{\mathbb{E}\left(\hat{h}_{\sf si}(h_{\sf si} - \hat{h}_{\sf
    si})(x_{\sf si}(t - \hat{\Delta}_{\sf si}) - x_{\sf si}(t -
  \Delta_{\sf si}))x_{\sf si}(t - \Delta_{\sf si})\right)\right\} + \sigma_{\sf noise}^2,
\label{eq:ResidueConventional}
 \end{eqnarray}
where (a) holds because of independence of thermal noise with
self-interference channel and the signal itself, (b) is true due to
assumption that the average power at the transmitter is unity. %% The
%% quality of channel estimate definitely dictates the amount the
%% cancellation.
%% In order to obtain the estimate of the
%% self-interference channel a training sequence can be used. The
%% training sequence is transmitted in a time slot when the intended
%% signal is not being transmitted, so that a cleaner estimate of the
%% self-interference channel is obtained. Let the discrete time training
%% sequence be a pseudo-noise sequence of length $T_{\sf train}$ denoted
%% by $\bar{x}_{\sf train} = [x_{\sf train}(1), \ldots, x_{\sf
%%     train}(T_{\sf train})]$ where $\mathbb{E}(|x_{\sf train}(i)|^2) =
%% 1, \forall i \in \{1, 2, \ldots, T_{\sf train}\}$.
Estimating a channel with single delay tap has been studied in
\cite{quazi}, where it is shown that estimation error in the channel
attenuation behaves as
\begin{equation}
\mathbb{E}\left(|\hat{h}_{\sf si} - h_{\sf si}|^2\right) = \frac{\sigma_{\sf
    noise}^2}{T_{\sf train}},
\label{eq:ErrorConventional}
\end{equation}
where $T_{\sf train}$ is the length of the training sequence used to
estimate the self-interference channel. Also, let $h_{\sf si, error}$
denote the error in the estimate of the channel attenuation, then
\begin{eqnarray}
  \mathbb{E}\left(\hat{h}_{\sf si}(h_{\sf si} - \hat{h}_{\sf si}))\right) &  = & \mathbb{E}\left((h_{\sf si, error} + h_{\sf si})h_{\sf si, error}\right)  =  \mathbb{E}\left((h_{\sf si, error})^2\right) + h_{\sf si}\mathbb{E}\left(h_{\sf si, error}\right)  =  \frac{\sigma_{\sf noise}^2}{T_{\sf train}}.
\label{eq:ErrorDelay}
\end{eqnarray}
In \cite{quazi}, it has been shown that the variance in the estimate
of the delay goes down as the inverse of training length $T_{\sf
  train}$. Moreover, it can be easily shown that for any bandlimited
signal $x_{\sf si}(t)$ and small enough $\Delta_{\sf si} -
\hat{\Delta}_{\sf si}$,
\begin{equation}
1 - R_{x_{\sf si}}(\Delta_{\sf si} - \hat{\Delta}_{\sf si}) \leq
c(\Delta_{\sf si} - \hat{\Delta}_{\sf si})^2,
\label{eq:BoundOnAutoCorr}
\end{equation}
where $R_{x_{\sf si}}(.)$ is the autocorrelation function of $x_{\sf
  si}(t)$ and $c$ is a positive constant (see Appendix
\ref{sec:Lowerbound} for details). Applying
\eqref{eq:ErrorConventional}, \eqref{eq:ErrorDelay},
\eqref{eq:BoundOnAutoCorr} and Equation (6) of \cite{quazi} to
\eqref{eq:ResidueConventional}, the residual self-interference for the
signal model in \eqref{eq:BasicModel} is bounded above as
\begin{equation}
  \sigma_{\sf residual}^2  < \frac{5\sigma_{\sf noise}^2}{T_{\sf train}} + \sigma_{\sf noise}^2,
\end{equation}
i.e., it decays inversely to the training length $T_{\sf
  train}$. Letting $T_{\sf train} \to \infty$ for
\eqref{eq:BasicModel}, the residual self-interference should only be
composed of thermal noise. Since the channel estimation error decays
inversely to the length of the training, for the signal model
described by \eqref{eq:BasicModel}, even with a very short training
length, say $T_{\sf train} = 5$ the residual self-interference is no
more than 3~dB above thermal noise. However, the observed phenomenon
in \cite{Duarte:phd} is that the residual self-interference is 15 dB
higher than the thermal noise which is clearly not explained by the
signal model in \eqref{eq:BasicModel}. In
\cite{Duarte:2010aa,Choi:2010aa,Choi:2011m} too the residual
self-interference is reported to be much higher than 15~dB above
thermal noise floor, thus we suspect that channel model in
\eqref{eq:BasicModel} does not capture all dominant sources of radio
impairments.

%% \subsection{Quantization Noise}
%% The signal to quantization noise ratio delivered by a 14-bit ADC is 84
%% dB {\bf Achal, why we talking about  14-bit ADC, this came out of blue.}. Thus the quantization noise is reasonably small in comparison to
%% the receiver thermal noise to neglect it as a source of error for
%% channel estimation. As a reference point, the experimental evaluation
%% in \cite{Duarte:phd} had a maximum signal to thermal noise of 50 dB at
%% the receiver, which is smaller than signal to quantization noise of 84
%% dB. If thermal noise cannot explain the bottleneck in active
%% cancellation, quantization noise too cannot. {\bf The last sentence makes it sound like that the two conclusions are linked. Are they really? If yes, not clear from the discussion.}

\section{Identifying the Bottleneck in Active Cancellation}
\label{sec:experiment}
 \subsection{Possible sources of bottleneck}
 Transmitter phase noise, receiver phase noise, IQ imbalance, power
 amplifier non-linearity and quantization noise are some of the other
 impairments in the transmit-receive chain at the full-duplex node
 which can possibly limit the amount of active analog cancellation. In
 \cite{Duarte:2010aa}, a 14-bit ADC is used, which delivers a signal
 to quantization noise ratio of 84~dB, making quantization noise much
 smaller than the thermal noise, thus ruling quantization noise out as
 a source of bottleneck in estimation of self-interference and
 consequently active analog cancellation. IQ imbalance does not vary
 significantly with time and can be easily calibrated, thus
 eliminating it as a source of bottleneck. Power amplifier shows
 significant non-linearity only when it is operated in its non-linear
 regime. In this paper, we want to explain the bottlenecks in current
 designs of full-duplex and since most of the designs to date have
 been designed in the linear regime of power amplifier, they do not
 suffer from power amplifier non-linearity.

\subsection{Experiment}
\begin{figure}[htb]
\centering
  \scalebox{0.6}{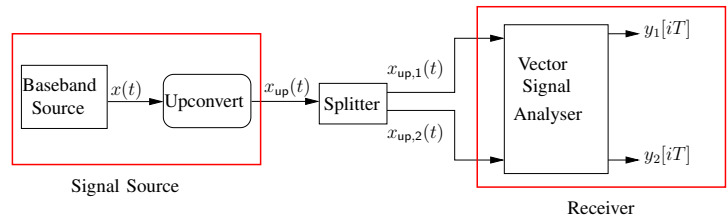}
  \caption{Schematic representation of the experiment in
    Section~\ref{sec:experiment} to acquire copies of a signals using
    a vector signal analyzer. WARP and Vector Signal Generator were
    two different signal sources considered in the experiment.}
    \label{fig:experiment}
\end{figure}

In our related work~\cite{asilomar2012}, we presented an experiment
through which we identify the bottleneck in active cancellation in a
full-duplex system. For the sake of completeness, we describe the
steps of the experiment and then explain how it is used to identify
the source of bottleneck in active cancellation. Following are the
steps of the experiment, schematically shown in the
Figure~\ref{fig:experiment}.
\begin{itemize}
  \item A signal $x(t) = e^{j\omega t}$ is digitally generated, with
    $\omega/2\pi = $1MHz, and is upconverted to the carrier frequency
    of $f_c = \omega_c/2\pi$. Let $x_{\sf up}(t)$ denote the
    upconverted signal.
  \item The signal $x_{\sf up}(t)$ is split using a 3-port power
    splitter \cite{pe2014}. Let $x_{\sf up,1}(t)$ and $x_{\sf up,
      2}(t)$ denote the two signals output from the power splitter.
  \item Using a wired connection, the signals $x_{\sf up,1}(t)$ and
    $x_{\sf up,2}(t)$ are fed into two input ports of a vector signal
    analyzer (VSA) \cite{VSA}. Using the knowledge of $\omega_c$, the
    VSA downconverts the received signals and digitizes them. Let the
    digitized signals, after downconversion be denoted by $y_1[iT]$
    and $y_2[iT]$. In the experiment $T$ was chosen to be 21.7 ns.
\end{itemize}
The above experiment is conducted using two signal sources: an
off-the-shelf radio chip \cite{maxim} used in WARP \cite{warp} and a
high precision Vector Signal Generator \cite{siggen}. For WARP $f_c =
\omega_c/2\pi = 2.4$~GHz and for the Vector Signal Generator $f_c =
\omega_c/2\pi = 2.2$~GHz.

\subsection{Mimicking active cancellation}
The received signal $y_1[iT]$ and $y_2[iT]$ are sequences of complex
numbers. To analyse the amount of cancellation, we treat $y_1[iT]$ as
the self-interference signal and use a processed version of $y_2[iT]$
as the cancelling signal. The transmitted signal is narrowband,
therefore if the upconversion process does not add any noise, then
\begin{eqnarray}
\label{eq:MimicCancellation1}  y_1[iT] & = & h_1e^{-j(\omega_c + \omega)\Delta_1}x[iT] + z_1[iT],  \\
\label{eq:MimicCancellation2} y_2[iT] & = & h_2e^{-j(\omega_c + \omega)\Delta_2}x[iT] + z_2[iT], 
\end{eqnarray}
where $h_1$ and $h_2$ are complex attenuations, and $\Delta_1$ and
$\Delta_2$ are delays introduced by the wires, and $z_1[iT]$ and
$z_2[iT]$ denote the thermal noise at the receiver. Using wires to
connect the source and receivers ensures that the temporal variation
in $h_1$ and $h_2$ is minimal. The wires are chosen of approximately
the same length so that $\Delta_1 \approx \Delta_2$. To mimic active
cancellation, we subtract a suitably scaled version of $y_2[iT]$ from
$y_1[iT]$, thereby leaving a residual self-interference which is given
by
\begin{equation}
  y_{\sf residual}[iT] = y_1[iT] - h_cy_2[iT],
\label{eq:ResidualInMimic}
\end{equation}
where $h_c$ is a complex number computed as
\begin{eqnarray}
 h_c & = &  \frac{\sum_{i = 1}^N y_2[iT]^{'}y_1[iT]}{\sum_{i = 1}^N |y_2[iT]|^2}.
    % & = & \frac{\sum_{i = 1}^N (h_2^{'}h_1x_{\sf up}(iT - \Delta_2)^{'}x_{\sf up}(iT - \Delta_1)) +h_2^*x_{\sf up}[iT - \Delta_2]^{*}e^{-j\omega_ciT}z_1(iT)}{\sum_{i = 1}^N}
\end{eqnarray}
Now consider a delayed version of the signal $y_2[iT]$,
\begin{eqnarray}
  y_2[(i - d)T] & = & h_2e^{-j(\omega_c + \omega)\Delta_2}x[(i- d)T] + z_2[(i-d)T] \nonumber \\
 & = & h_2e^{-j((\omega_c + \omega)\Delta_2 + \omega d T)}x[iT] + z_2[(i -d)T],
\end{eqnarray}
where $d$ is a non-negative integer. We can subtract a scaled version
of $y_2[(i - d)T]$ from $y_1[iT]$ such that the residual
self-interference is
\begin{equation}
  y_{\mathsf{residual}, d}[iT] = y_1[iT] - h_c(d)y_2[(i - d)T],
\end{equation}
where the scaling $h_c(d)$ is computed as 

\begin{equation}
  h_c(d)  =   \frac{\sum_{i = 1}^N y_2[(i - d)T]^{'}y_1[iT]}{\sum_{i = 1}^N |y_2[(i -d)T]|^2}.
\label{eq:CorrectionWithDelay}
\end{equation}
If \eqref{eq:MimicCancellation1} and \eqref{eq:MimicCancellation2}
hold true and if we rewrite $h_c(d) =
\frac{h_1}{h_2}e^{j(\omega_c(\Delta_2 - \Delta_1) + \omega d T)} +
\epsilon$, then the expected strength of the residual signal is given
by
\begin{eqnarray}
  \mathbb{E}(|y_{\mathsf{residual}, d}[iT]|^2)  & = & \mathbb{E}(|y_1[iT] - h_c(d)y_2[(i -d)T]|^2) \nonumber \\
 & = & |h_2|^2\mathbb{E}(|\epsilon|^2) + \mathbb{E}(|z_1[iT]|^2 + |z_2[(i -d)T]|^2) = |h_2|^2\mathbb{E}(|\epsilon|^2) + 2\sigma_{\sf noise}^2.
\end{eqnarray}
In Appendix~\ref{sec:DelayScaling}, we show that by letting $N \to
\infty$ we have
\begin{equation}
   \mathbb{E}(|y_{\mathsf{residual}, d}[iT]|^2) = \frac{|h_1|^2}{|h_2|^2}\sigma_{\sf noise}^2 + 2\sigma_{\sf noise}^2.
\end{equation}
 For the experiment conducted $\frac{|h_1|^2}{|h_2|^2} \approx 1$,
 thus the strength of the residual self-interference should be
 approximately $3\sigma_{\sf noise}^2$. The analysis reveals that if
 \eqref{eq:MimicCancellation1} and \eqref{eq:MimicCancellation2} hold
 true, then the amount of cancellation should be independent of the
 delay $d$ and dependent only on the thermal noise.

\subsection{Experiment: Results and their explanation}
\label{sec:exp-results}
\begin{figure}[htb]
\begin{minipage}[b]{0.45\linewidth}
\centering
\scalebox{0.35}{\includegraphics{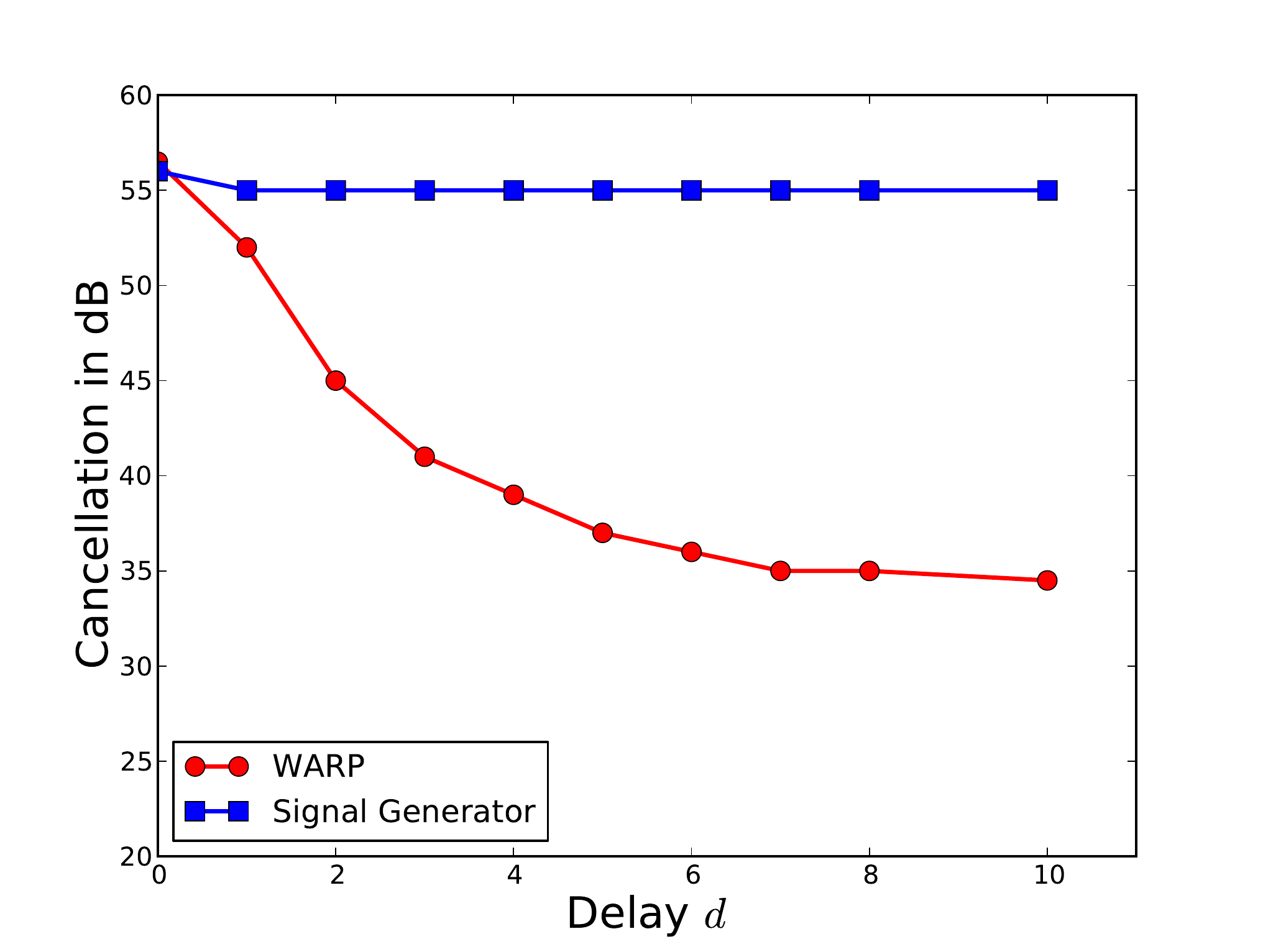}}
 \caption{Amount of cancellation as a function of the delay for
   different signal sources measured from the experiment in
   Section~\ref{sec:experiment}. Also shown in our related
   work~\cite{asilomar2012}.}
  \label{fig:CancelvsDelay}
\end{minipage}
\hspace{0.5cm}
\begin{minipage}[b]{0.45\linewidth}
\centering
\scalebox{0.35}{\includegraphics{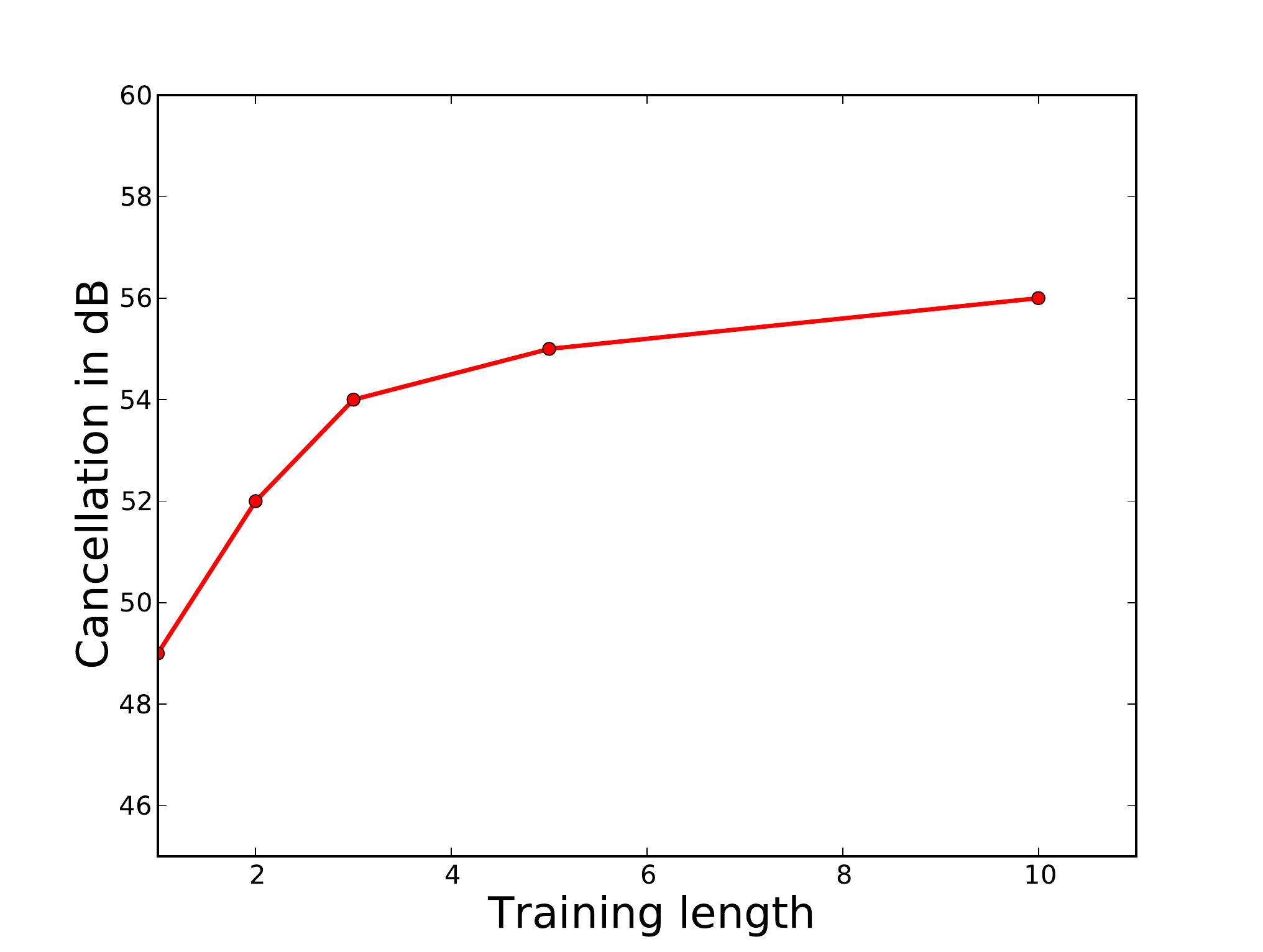}}
  \caption{Amount of active cancellation as a function of the training
    length for a delay $d = 0$ for WARP as the signal source measured
    from the experiment in Section~\ref{sec:experiment}.}
  \label{fig:CancelvsTraining}
\end{minipage}
\end{figure}

In Figure~\ref{fig:CancelvsDelay}, we plot the amount of cancellation
as a function of delay $d$ measured from the experiment for both the
signal sources. For WARP as the signal source, when $d$ is small then
the amount of cancellation depends on the delay. As the delay
increases the cancellation floors around 35~dB. The measurement from
the experiment shows that for WARP as a signal source, even for a
delay $d=100$, the amount of cancellation is approximately 35~dB. On
the other hand, the amount of cancellation when the vector signal
generator is used as a signal source is approximately 55~dB,
independent of the delay.
%% \begin{figure}[!h]
%% %  \centering
%%   \scalebox{0.5}{\includegraphics{figs/plot-cancelvsdelay.pdf}}
%%   \caption{Amount of cancellation as a function of the delay for
%%     different signal sources measured from the experiment}
%%   \label{fig:CancelvsDelay}
%% %\includegraphics{figs/plot-cancelvsdelay.pdf}
%% \end{figure}

\paragraph{Upper bound of cancellation} 
For both signal sources, the upper bound of cancellation is around
55~dB. The limitation on the cancellation can be explained by the
dynamic range of the measurement equipment. The data-sheet
\cite{VSA} of the VSA lists that it offers a dynamic range of
anywhere between 55-60~dB. Thus, the received signals $y_1[iT]$ and
$y_2[iT]$ themselves have an SNR of no more 55-60~dB, thereby limiting
the maximum cancellation in the range of 55-60~dB only.

\paragraph{Phase noise explains the trend of cancellation} 
Two observations from the experiment conducted, when WARP is used as a
signal source source, need an explanation. The first observation is
that the amount of cancellation reduces by increasing the delay
between self-interference signal and cancelling signal. And the second
observation is that the amount of cancellation has a lower bound of
$\approx$35~dB. We claim that both the observations can be explained
if we consider the perturbations introduced by phase noise in the
upconverted signal.

Phase noise is the jitter in the local oscillator. If the baseband
signal $x(t)$ is upconverted to a carrier frequency of $\omega_c$,
then the upconverted signal $x_{\sf up}(t) = x(t)e^{j(\omega_c t +
  \phi(t))}$, where $\phi(t)$ represents the phase noise. While
downconverting a signal, phase noise can be similarly defined. The
variance of phase noise is defined as $\sigma_{\phi}^2 =
\mathbb{E}(|\phi(t)|^2)$ and its autocorrelation function is denoted
by $R_{\phi}(.)$. For a measurement equipment like VSA, the phase
noise at the receiver is small. Therefore the total phase noise in the
received signal, after downconversion, is dominated by phase noise at
transmitter, i.e., the source of the signal. In presence of phase
noise, the equations \eqref{eq:MimicCancellation1} and
\eqref{eq:MimicCancellation2} can be rewritten as
\begin{eqnarray}
\label{eq:PhaseNoise1}  y_1[iT] & = & h_1e^{-j(\omega_c + \omega)\Delta_1}e^{j\phi[iT - \Delta_1]}x[iT] + z_1[iT],  \\
\label{eq:PhaseNoise2} y_2[iT] & = & h_2e^{-j(\omega_c + \omega)\Delta_2}e^{j\phi[iT - \Delta_2]}x[iT] + z_2[iT]  .
\end{eqnarray}
For a delay $d$, suppose an oracle provides scaling $h(d) =
\frac{h_1}{h_2}e^{j(\omega(\Delta_2 - \Delta_1) + \omega d T)}$ to
subtract a delayed version of $y_2[iT]$ from $y_1[iT]$, then the
residual self-interference will be given by 
\begin{eqnarray}
  y_{\mathsf{residual}, d} [iT] & = & y_1[iT] - h(d)y_2[(i - d)T]
  \nonumber \\ & = & h_1 x[iT]e^{-j(\omega_c +
    \omega)\Delta_1}(e^{j\phi[iT - \Delta_1]} - e^{j\phi[iT - \Delta_2
      - dT]}) + z_1[iT] - z_2[(i-d)T] \nonumber \\ &
  \stackrel{\text{(a)}}{\approx} & jh_1 x[iT]e^{-j(\omega_c +
    \omega)\Delta_1}(\phi[iT - \Delta_1] - \phi[iT - \Delta_2 - dT]) +
  z_1[iT] - z_2[(i-d)T], \nonumber
\end{eqnarray}
where (a) is valid if the phase noise is small. The resulting strength
of the residual self-interference is
\begin{eqnarray}
  \mathbb{E}(|y_{\mathsf{residual}, d}[iT]|^2) & \approx & |h_1|^2\sigma_{\phi}^2(1 - R_{\phi}(\Delta_2 - \Delta_1 + dT)) + 2\sigma_{\sf noise}^2 \nonumber \\
& \stackrel{\text{(a)}}{\approx} & |h_1|^2\sigma_{\phi}^2(1 - R_{\phi}(dT)) + 2\sigma_{\sf noise}^2. \label{eq:residue}
\end{eqnarray}
In \eqref{eq:residue}, the approximation (a) is reasonable since
$\Delta_1 \approx \Delta_2$. In the absence of phase noise, using
$h(d)$ as the scaling for cancellation leads to a residual
self-interference dependent only on thermal noise. In presence of
phase noise, the strength of the residual self-interference is a
function of the delay $d$. As the delay increases, it is natural that
the temporal correlation in phase noise reduces. Therefore the amount
of cancellation, when WARP is used as a signal source, will reduce as
the delay increases which explains the trend of cancellation in
Figure~\ref{fig:CancelvsDelay}. Once the delay is sufficiently large,
the residual self-interference depends only on the variance of the
phase noise and thermal noise. For the MAXIM 2829 transceiver used in
WARP, $\sigma_{\phi} \approx 0.7^\circ$ (see
Appendix~\ref{sec:VarPhaseNoise} for calculations), which is
equivalent to 35~dB cancellation for large delay $d$ which explains
lower bound of cancellation. Although the trend in cancellation when
signal generator is used as the source does not appear to be similar
to WARP, it can be explained using its phase noise figure. At 2.2~GHz,
the vector signal generator \cite{siggen} has a phase noise variance
given by $\sigma_{\phi} = 0.06^\circ$. The corresponding lower bound
of the cancellation is $\approx$55 dB. Thus, the lower bound due to
phase noise is close to upper bound of cancellation due to dynamic
range limitations of the VSA, thereby showing no apparent variation of
cancellation with delay.

\noindent \paragraph{Impact of estimation error} To strengthen our
argument that phase noise is the dominant source of bottleneck in the
cancellation in the experiment and not estimation error, we plot the
amount of cancellation measured as function of the number of training
samples used to obtain $h_c = h_c(0)$ in
Figure~\ref{fig:CancelvsTraining}. Reducing the number of training
samples will increase the error in estimation of $h_c(0)$. Figure
\ref{fig:CancelvsTraining} shows that in the controlled experiment,
reducing the number of training samples to estimate $h_c(0)$ reduces
the amount of cancellation by no more than 6~dB for the WARP as the
signal source. Phase noise can explain the variation in cancellation
of 20 dB observed and plotted in Figure~\ref{fig:CancelvsDelay} for
varying delays, while estimation error can explain at-most 6 dB of
variation, therefore phase noise is the dominant source of bottleneck
in active cancellation.

\section{Answer 1. Impact of Phase Noise on Active Analog Cancellation}
\label{sec:AllCancellers}
In this section, we answer ``What limits the amount of active analog
cancellation in a full-duplex system design?'' We quantify the impact
of transmitter and receiver phase noise on the amount of active analog
cancellation achieved by different types of active analog cancellers
described in Section~\ref{sec:ActiveAnalogCanceller}.

A quick note on the notation for the subsequent discussion.  Phase
noise and its corresponding variance in the self-interference path and
cancelling path are denoted by the pairs $(\phi_{\sf si}(t),
\sigma_{\sf si}^2)$ and $(\phi_{\sf cancel}(t), \sigma_{\sf
  cancel}^2)$ respectively, while the phase noise at the receiver and
its variance is denoted by the pair $(\phi_{\sf down}(t),\sigma_{\sf
  down}^2)$. For simplicity of analysis, we assume that the phase
noise at the transmitter, $\phi_{\sf si}(t)$ and $\phi_{\sf
  cancel}(t)$, are independent of the phase noise at the receiver,
$\phi_{\sf down}(t)$. 
\subsection{Impact of phase noise on pre-mixer cancellers}
\label{sec:Pre-radioPerfect}

\noindent \emph{Result 1 \cite{asilomar2012}: The amount of active
  analog cancellation in pre-mixer cancellers is limited by the
  inverse of the variance of phase noise. Moreover, matching local
  oscillators in the self-interference and cancelling paths can
  increase the amount of active analog cancellation.}

To highlight the impact of transmitter phase noise, we first analyse a
special scenario for pre-mixer analog cancellers when the
self-interference channel, $\mathbf{h}_{\sf si}(t)$, is perfectly
known to the canceller. The self-interference channel is
$\mathbf{h}_{\sf si}(t) = h_{\sf si}\delta(t - \Delta_{\sf si})$,
therefore the cancelling signal prior to upconversion, designed by
exploiting the knowledge of the self-interference is
\begin{equation}
  x_{\sf cancel}(t) = -h_{\sf si}x(t - \Delta_{\sf si})e^{-j\omega_c\Delta_{\sf si}}.
\label{eq:CancellingSignal}
\end{equation}
It is easy to verify that in the absence of any phase noise, the
cancelling signal in \eqref{eq:CancellingSignal} will null
the self-interference signal at the receiver. In presence of phase
noise, the cancelling signal after upconversion will be $x_{\sf
  cancel}(t)e^{j(\omega_ct + \phi_{\sf cancel}(t))}$. At the receiver,
the self-interference and the cancelling signal add up, which upon
downconversion result in the following residual self-interference signal
\begin{eqnarray}
  y_{\sf residue-analog}(t) & = & \left(h_{\sf si}x_{\sf si}(t -
  \Delta_{\sf si})e^{-j\omega_c\Delta_{\sf si}}e^{j\phi_{\sf si}(t -
    \Delta_{\sf si})} - h_{\sf si}x_{\sf si}(t - \Delta_{\sf
    si})e^{-j\omega_c\Delta_{\sf si}}e^{j\phi_{\sf
      cancel}(t)}\right)e^{-j\phi_{\sf down}(t)} \nonumber \\ && + z_{\sf noise}(t)
  \nonumber \\ 
& = & h_{\sf si}x(t - \Delta_{\sf
    si})e^{-j\omega_c\Delta_{\sf si}}\left(e^{j\phi_{\sf si}(t -
    \Delta_{\sf si})} - e^{j\phi_{\sf cancel}(t)}\right)e^{-j\phi_{\sf down}(t)} + z_{\sf noise}(t).
\label{eq:ResidueAnalog}
\end{eqnarray}
Equation~\eqref{eq:ResidueAnalog} assumes that the upconverting and
downconverting frequencies are identical, which is valid since both
the upconvertor and downconvertor are on the same node. Assuming that
the magnitude of phase noise is small, the residual self-interference
can be approximated as
\begin{equation}
  y_{\sf residue-analog}(t) \approx  h_{\sf si}x(t - \Delta_{\sf si})e^{-j\omega_c\Delta_{\sf si}}e^{-j\phi_{\sf down}(t)}\left(j\phi_{\sf si}(t - \Delta_{\sf si}) - j\phi_{\sf cancel}(t)\right) + z_{\sf noise}(t),
\label{eq:ApproxResidueAnalog}
\end{equation}
and the power of the residual self-interference is computed as
\begin{eqnarray}
  \mathbb{E}(|y_{\sf residue-analog}(t)|^2) &
  \stackrel{\text{(a)}}{\approx} & |h_{\sf si}|^2\mathbb{E}(|x(t -
  \Delta_{\sf si})|^2)|e^{-j\omega_c\Delta_{\sf si}}e^{-j\phi_{\sf
      down}(t)}|\mathbb{E}(|\phi_{\sf si}(t - \Delta_{\sf si}) -
  \phi_{\sf cancel}(t)|^2) + \sigma_{\sf noise}^2 \nonumber \\ &
  \stackrel{\text{(b)}}{=} & |h_{\sf si}|^2\mathbb{E}(|\phi_{\sf si}(t
  - \Delta_{\sf si}) - \phi_{\sf cancel}(t)|^2) + \sigma_{\sf noise}^2
  ,\label{eq:ApproxResidueAnalogStrength}
%& \stackrel{\text{c}}{=} & |h_{\sf si}|^2(\sigma_{\sf si}^2 +\sigma_{\sf cancel}^2) + \sigma_{\sf noise}^2,
\end{eqnarray}
where (a) holds since the thermal noise is independent of the
self-interference and phase noise, (b) holds because of the unit power
constraint at the transmitter. Now, we elaborate the observations on
\emph{Result~1} based on \eqref{eq:ApproxResidueAnalogStrength} which
were briefly highlighted in our related work~\cite{asilomar2012}.

\begin{figure}[htb]
  \centering
  \scalebox{0.5}{\includegraphics{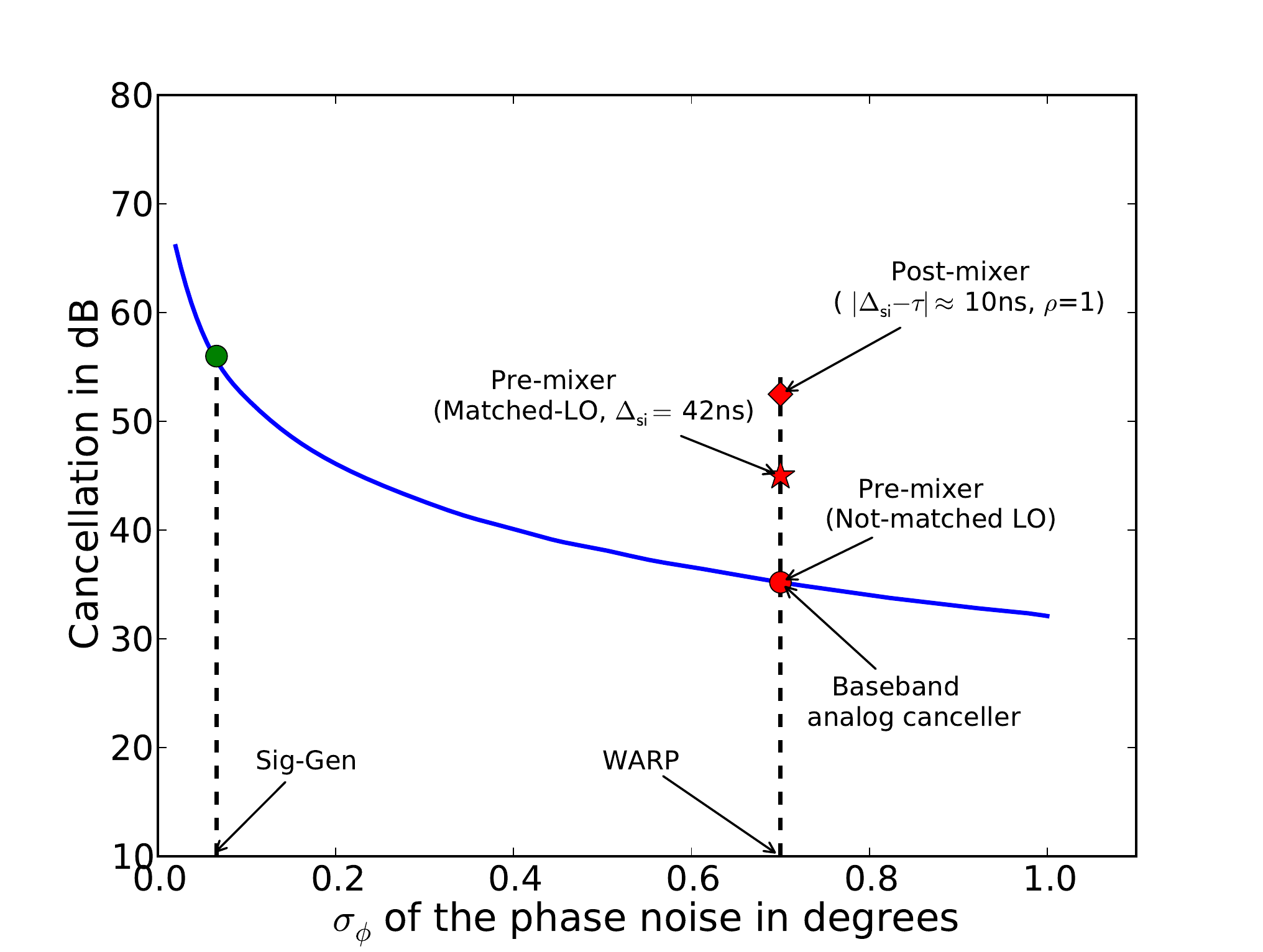}}
  \caption{Amount of active analog cancellation possible in different
    types of cancellers as function of phase noise. The solid curve is
    a plot of amount of cancellation possible in pre-mixer cancellers
    if LOs are not matched, as a function of the variance of phase noise.}
  \label{fig:CancelvsPhi}
\end{figure}

\emph{Observation 1}:
If the local oscillators supplied to the self-interference path
  and the cancelling path are different, as is the case in
  \cite{Duarte:2010aa}, then the correlation between the $\phi_{\sf
    si}(t)$ and $\phi_{\sf cancel}(t)$ is zero. With the assumption
  that $\sigma_{\sf si}^2 = \sigma_{\sf cancel}^2$, the strength of
  the residual self-interference is
  \begin{equation}
    \mathbb{E}(|y_{\sf residue-analog}(t)|^2) \approx 2|h_{\sf si}|^2\sigma_{\sf si}^2  + \sigma_{\sf noise}^2.
    \label{eq:ResidueStrengthPreRadio} 
  \end{equation}
  Note that the strength of the self-interference before active analog
  cancellation is $|h_{\sf si}|^2$. Therefore
  \eqref{eq:ResidueStrengthPreRadio} implies that the strength of
  residual self-interference after active analog cancellation is
  dependent on the strength of the self-interference before
  cancellation. The amount of active cancellation is given by
  $\frac{|h_{\sf si}|^2}{2|h_{\sf si}|^2\sigma_{\sf si}^2 +
    \sigma_{\sf noise}^2} \leq \frac{1}{2\sigma_{\sf si}^2}$. Thus,
  $\frac{1}{2\sigma_{\sf si}^2}$ is an upper bound for the amount of
  cancellation in pre-mixer cancellers where the local oscillators in
  self-interference path and cancelling path are independent, which we
  plot in Figure~\ref{fig:CancelvsPhi}. Since \cite{Duarte:2010aa} is
  a pre-mixer canceller and is designed on WARP platform, where local
  oscillators in the cancelling and self-interference path are not
  matched, Figure~\ref{fig:CancelvsPhi} predicts the amount of active
  analog cancellation to be 35~dB which is very close to the amount of
  cancellation reported by \cite{Duarte:2010aa}.
 
\emph{Observation 2}: If the local oscillators in the self-interference path and the
  cancelling path are matched, $\phi_{\sf si}(t) = \phi_{\sf
    cancel}(t)$, then we have
  \begin{equation}
    \mathbb{E}(|y_{\sf residue-analog}(t)|^2) \approx 2|h_{\sf si}|^2\sigma_{\sf si}^2 (1 - R_{\phi_{\sf si}}(\Delta_{\sf si})) + \sigma_{\sf noise}^2.
    \label{eq:ResidueStrengthPreRadioa}
  \end{equation}
Equation~\eqref{eq:ResidueStrengthPreRadioa} indicates that for a
small delay $\Delta_{\sf si}$, the measure of the time of flight of
the self-interference signal, the temporal correlation of phase noise
aids in reducing the residual self-interference in pre-mixer
cancellers. In Section~\ref{sec:exp-results}, we measured and plotted
in Figure~\ref{fig:CancelvsDelay}, the amount of active analog
cancellation as a function of the delay $\Delta_{\sf si}$, for a
narrowband signal source. For $\Delta_{\sf si}\approx 42$ns, the time
of flight of self-interference signal for 12 meters, the measurements
in Figure~\ref{fig:CancelvsDelay} tell us that matching local
oscillators in the self-interference and cancelling path will yield an
active analog cancellation of 45~dB. Thus, matching local oscillators,
when WARP is used as a signal source, results in 10~dB higher active
analog cancellation compared to when local oscillators are not
matched. In \cite{Duarte:2010aa}, ergodic rate of full-duplex beats
half-duplex only upto 3.5 meters (indoor). However,
in~\cite{Duarte:phd}, an additional 10~dB passive suppression results
in higher ergodic rates for half-duplex upto 6 meters. Matching local
oscillators in \cite{Duarte:phd} will give another 10~dB increase in
overall reduction making full-duplex attractive at reasonable WiFi
ranges.

From \eqref{eq:ResidueStrengthPreRadio} and
\eqref{eq:ResidueStrengthPreRadioa}, we know that the phase noise
dependent residual scales linearly in strength with
self-interference. Therefore at higher received self-interference
powers, phase noise becomes the dominant source of residual
self-interference after active analog cancellation in pre-mixer
cancellers.

\subsection{Performance of different active analog cancellers with imperfect channel estimates}
\label{sec:AllCancelersImperfect}
Now we analyze and compare the impact of phase noise on active analog
cancellation in pre-mixer, post-mixer and baseband analog
cancellers. In order to draw the comparison, we analyse the amount of
active analog cancellation when the estimate of self-interference
channel is imperfect for which we show the following:

\noindent \emph{Result 2: For pre-mixer, post-mixer, as well as
  baseband analog canceller, the amount of active cancellation is
  inversely proportional to the variance of phase noise. However, the
  constant of proportionality is different for each canceller leading
  to different amounts of active analog cancellation.}

To model imperfection, we let $\widehat{\mathbf{h}}_{\sf si}(t) = \rho
h_{\sf si}\delta(t - \tau)$ denote the imperfect channel estimate of
the self-interference channel, where $(1 - \rho)$ and $(\tau -
\Delta_{\sf si})$ represent the error in estimate of channel
attenuation and delay respectively. Setting $\rho = 1$ and $\tau =
\Delta_{\sf si}$, we obtain the special case of perfect channel
estimates.

The objective of each of the cancellers is to create a perfect null
for the self-interference signal. However, in presence of phase noise
each canceller adds a slightly different cancelling signal to the
self-interference signal. Based on the imperfect channel estimate, the
canceller generates $-\rho h_{\sf si} x_{\sf si}(t -
\tau)e^{-j\omega_c\tau}$ as the cancelling signal. The cancelling
signal after downconversion at the receiver will appear in analog
baseband as
\begin{equation}
  x_{\sf cancel, pre}(t) = -\rho h_{\sf si} e^{j(-\omega_c \tau +
    \phi_{\sf cancel}(t) - \phi_{\sf down}(t))} h_{\sf si}x_{\sf si}(t
  - \tau).
\label{eq:PreRadioCancelSig}
\end{equation}
Note that the cancelling signal in pre-mixer analog cancellers is
actually added to the received signal at RF, and then the combined
signal is downconverted. However, in \eqref{eq:PreRadioCancelSig} we
explicitly show the contribution of the cancelling signal in the
residual self-interference signal after downconversion.

For the post-mixer analog canceller, the equivalent of
\eqref{eq:PreRadioCancelSig} can be written as
\begin{equation}
  x_{\sf cancel, post}(t) = -\rho e^{j(-\omega_c \tau + \phi_{\sf cancel}(t - \tau) - \phi_{\sf down}(t))} h_{\sf si}x_{\sf si}(t  - \tau).
\label{eq:PostRadioCancelSig}
\end{equation}
Note that \eqref{eq:PreRadioCancelSig} and
\eqref{eq:PostRadioCancelSig} differ in the amount of delay the
transmitter phase noise encounters. We remind the reader that in
post-mixer analog cancellers, the cancelling signal is identical to
the transmitted signal until after upconversion and therefore the
phase noise $\phi_{\sf cancel}(t) = \phi_{\sf si}(t)$.

Finally, in baseband analog cancellers the cancelling signal has the
following contribution to the residual self-interference
\begin{equation}
  x_{\sf cancel, bb}(t) = -\rho e^{-j\omega_c \tau} h_{\sf si}x_{\sf si}(t - \tau).
\label{eq:BasebandRadioCancelSig}
\end{equation}
The cancelling signal in \eqref{eq:BasebandRadioCancelSig} is not
perturbed by any phase noise because the cancelling signal does not go
through the RF chain itself.
% {\bf Achal, this is poor transition. The sentence before is baseband and then you switch to pre-radio} 

Having described the cancelling signal, we can now write the residual
self-interference for pre-mixer, post-mixer and baseband analog
cancellers by adding the cancelling signal to the self-interference
signal at the receiver. The residual self-interference for pre-mixer
cancellers is
\begin{eqnarray}
y_{\sf residual-analog}(t) = e^{-j(\omega_c\Delta_{\sf
    si} + \phi_{\sf si}(t - \Delta_{\sf si}) - \phi_{\sf
    down}(t))}h_{\sf si}x_{\sf si}(t - \Delta_{\sf si}) + x_{\sf
  cancel, pre}(t) + z_{\sf noise}(t).
\label{eq:ResiduePreRadioRep}
\end{eqnarray}
The residual self-interference for post-mixer and baseband analog
cancellers is defined similar to \eqref{eq:ResiduePreRadioRep}, by
substituting the appropriate cancelling signal from
\eqref{eq:PostRadioCancelSig} and
\eqref{eq:BasebandRadioCancelSig}. 

We are interested in the strength of the residual self-interference
after analog cancellation, and a close approximation can be found
making use of the assumption that $\phi_{\sf si}(t) << 1, \phi_{\sf
  cancel}(t) << 1, \phi_{\sf down}(t) << 1$. The computation is shown
in the Appendix~\ref{sec:ResidualComputations} and the resulting
strength of the residual self-interference is listed in
Table~\ref{table:ResidualEnergy}. From
Table~\ref{table:ResidualEnergy}, we make the following important
observations.

\emph{Observation 3:} Due to imperfect channel estimates, the
 strength of the residual self-interference in all the cancellers is
 composed of two types of residuals. The first type of residual
 self-interference is dependent only on the self-interference signal
 and the second type is dependent on phase noise. For all cancellers,
 the first type of residual self-interference dependent only on the
 self-interference signal is $|h_{\sf si}|^2(1 + |\rho|^2 - 2|\rho|
 R_{x_{\sf si}}(\Delta_{\sf si} - \tau))$ which vanishes if $\rho = 1$
 and $\tau = \Delta_{\sf si}$, i.e., when perfect channel estimate is
 available. The second type of residual self-interference, dependent
 upon phase noise, scales with the variance of phase noise, as well as
 the strength of the self-interference channel $|h_{\sf si}|^2$, for
 all the cancellers. Due to the second type of residual
 self-interference linearly scaling with the variance of phase noise,
 the amount of active analog cancellation in the pre-mixer, post-mixer
 and baseband analog cancellers depend on the inverse of the variance
 of phase noise.

\emph{Observation 4:} In post-mixer cancellers, the strength of
residual self-interference due to phase noise is scaled by $(1 -
R_{\phi_{\sf si}}(\Delta_{\sf si} - \tau))$. The autocorrelation
function $R_{\phi_{\sf si}}(.)$ approaches unity as the error in
estimating the delay of the channel, $(\Delta_{\sf si} - \tau)$, is
reduced, thereby reducing the residual self-interference. Unlike
pre-mixer cancellers, where the delay $\Delta_{\sf si}$ determines the
amount of residual self-interference, post-mixer cancellers can reduce
residual self-interference by reducing the error in estimate of
self-interference channel. Figure~\ref{fig:CancelvsPhi} shows the
representative amount of cancellation of a post-mixer canceller for a
narrowband signal source where $|\Delta_{\sf si} - \tau| \approx 10$ns
and $\rho = 1$. In principle, higher cancellation in post-mixer
cancellers, as observed in \cite{Choi:2011m,Khojastepour:2011} is
possible, because unlike pre-mixer cancellers, the residual
self-interference continues to decrease as the error in the estimate
of self-interference channel improves. In \cite{Choi:2010aa} USRP
radios are used, whose phase noise variance (although not reported) is
likely to be higher than WARP radios, thus explaining low, 20 dB,
active analog cancellation.

\emph{Observation 5:} In baseband analog cancellers, the residual
self-interference scales as the sum of the variance of phase noise at
the transmitter and the receiver. Due the asummption that phase noise
in the local oscillator in the upconverting and downconverting circuit
are independent, baseband analog cancellers have a phase noise
dependent residual self-interference which does not depend on the
delay $\Delta_{\sf si}$. Even when $\rho = 1, \Delta_{\sf si} = \tau$,
amount of cancellation is upper bounded by $\frac{1}{\sigma_{\sf si}^2
  + \sigma_{\sf down}^2}$, which is similar to the performance of
pre-mixer cancellers with independent mixers in cancelling and
self-interference path as shown in Figure~\ref{fig:CancelvsPhi}.

\section{Answer 2. Benefit of Digital Cancellation after Active Analog Cancellation} 
\label{sec:DigitalCancellation}
In this section, we answer ``How do the amounts of cancellations by
active analog and digital cancelers depend on each other in a cascaded
system?''
\subsection{Digital cancellation when active analog cancellation uses perfect channel estimate}
\emph{Result 3: If active analog cancellation is performed with perfect channel estimates, then
\begin{itemize}
\item Digital cancellation does not reduce the strength of the
  residual self-interference at all, if $\phi_{\sf si}(t)$ and
  $\phi_{\sf cancel}(t)$ are identically distributed in pre- and
  post-mixer cancellers, and $\phi_{\sf si}(t)$ and $\phi_{\sf
    down}(t)$ are identically distributed in baseband analog
  cancellers.
\item If $\phi_{\sf si}(t)$ and $\phi_{\sf cancel}(t)$ are not
  identically distributed, then under the assumption that $\phi_{\sf
    si}(t) << 1$, $\phi_{\sf cancel}(t) << 1$, $\phi_{\sf down}(t)
  <<1$, digital cancellation does not help.
\end{itemize}}

For pre-mixer cancellers, the above result was already shown in our
related work~\cite{asilomar2012}. Digital cancellation can reduce the
residual only if $y_{\sf resdiue-analog}[iT]$ is correlated with the
self-interference signal $x_{\sf si}[iT]$. When $y_{\sf
  resdiue-analog}[iT]$ is correlated with $x_{\sf si}[iT]$, then
digital cancellation can reduce the strength of the residual
self-interference by subtracting a function of $x_{\sf si}[iT]$ from
$y_{\sf residue-analog}[iT]$. We consider the residual after active
analog cancellation in a pre-mixer canceller, as an example to show
that $y_{\sf resdiue-analog}[iT]$ is not correlated with $x_{\sf
  si}[iT]$. The correlation of the residual signal with $x_{\sf
  si}[iT]$ yields
\begin{eqnarray}
& & \mathbb{E}(y_{\sf residue-analog}[iT]x_{\sf si}[iT]) \nonumber \\ & = & \mathbb{E}(y_{\sf
  residual-si}[iT]x_{\sf si}[iT]) + \mathbb{E}(z_{\sf noise}[iT]x_{\sf
  si}[iT]) \nonumber \\
 & \stackrel{\text{(a)}}{=} & \mathbb{E}\left(y_{\sf
 residual-si}[iT]x_{\sf si}[iT]\right) \nonumber \\ 
& \stackrel{\text{(b)}}{=} & h_{\sf si}\mathbb{E}\left(x_{\sf
  si}[iT]x_{\sf si}[iT - \Delta_{\sf si}](e^{j\phi_{\sf si}[iT]} -
e^{j\phi_{\sf cancel}[iT - \Delta_{\sf si}]})e^{-j\phi_{\sf down}[iT]}\right)e^{-j\omega_c\Delta_{\sf si}}
\nonumber \\ & \stackrel{\text{(c)}}{=} & h_{\sf si}R_{x_{\sf
    si}}(\Delta_{\sf si})\mathbb{E}\left(e^{j\phi_{\sf si}[iT]} - e^{j\phi_{\sf
  cancel}[iT - \Delta_{\sf si}]}\right)\mathbb{E}\left(e^{-j\phi_{\sf down}[iT]}\right)
\label{eq:digitaldoesnothelp}
\end{eqnarray}
where $y_{\sf residual-si}[iT]$ denotes the residual
self-interference, in a pre-mixer canceller, minus thermal noise. In
equation \eqref{eq:digitaldoesnothelp}, equality (a) is true because
the thermal noise is zero mean and independent of the
self-interference, (b) is due to \eqref{eq:ResidueAnalog}, (c) holds
because phase noise is independent of the self-interference signal.
Suppose that $\phi_{\sf si}(t)$ and $\phi_{\sf cancel}(t)$ are
identically distributed, then $\mathbb{E}\left(e^{j\phi_{\sf si}[iT]}
- e^{j\phi_{\sf cancel}[iT - \Delta_{\sf si}]}\right) = 0$ letting us extend
\eqref{eq:digitaldoesnothelp} to
\begin{equation}
  \mathbb{E}(y_{\sf residue-analog}[iT]x_{\sf si}[iT]) = 0.
\label{eq:digitaldoesnothelp2}
\end{equation}
%% Since the active analog canceller used perfect estimate of the
%% channel, the residual signal is naturally composed of only noise,
%% therefore its correlation with $x_{\sf si}(t)$ is zero. Thus, given
%% that the active analog canceller was implemented with perfect estimate
%% of the self-interference channel, the digital canceller will not
%% reduce the self-interference any further.

%% Under the apprxomation $\phi_{\sf si}(t) << 1$, $\phi_{\sf cancel}(t)
%% << 1$, we can use the approximation in \eqref{eq:ApproxResidueAnalog}
%% to rewrite \eqref{eq:digitaldoesnothelp} as
%% \begin{equation}
%%   \mathbb{E}(y_{\sf residue-analog}[iT]x_{\sf si}[iT]) 
%% \approx
%% h_{\sf si}R_{x_{\sf si}}(\Delta_{\sf si})\mathbb{E}\left(j\phi_{\sf si}[iT] - j\phi_{\sf cancel}[iT - \Delta_{\sf si}]\right)
%% \mathbb{E}\left(e^{-j\phi_{\sf down}[iT]}\right) \stackrel{\text{(a)}}{=}  0
%% \label{eq:digitaldoesnothelp3}
%% \end{equation}
%% where (a) is true because phase noise is assumed to be zero
%% mean. Note Equation \eqref{eq:digitaldoesnothelp3} tells us that the
%% correlation of the residual self-interference with the
%% self-interference signal is approximately zero, thus digital
%% cancellation cannot reduce the strength of residual self-interference
%% any further.

%% test writing 

Under the approximation $\phi_{\sf si}(t) << 1$, $\phi_{\sf cancel}(t)
<< 1$, the residual self-interference signal in pre-radio cancellers
is given by \eqref{eq:ApproxResidueAnalog}. From
\eqref{eq:ApproxResidueAnalog}, we know that the residual
self-interference has a component where the signal, $x_{\sf si}(t -
\Delta_{\sf si})$, is multiplied by $j(\phi_{\sf si}(t -\Delta_{\sf
  si}) - \phi_{\sf si}(t))$. The difference of phase noises,
$j(\phi_{\sf si}(t -\Delta_{\sf si}) - \phi_{\sf si}(t))$, is zero
mean, independent of the signal, $x_{\sf si}(t - \Delta_{\sf si})$,
and changes every sample. Thus, the residual self-interference in
\eqref{eq:ApproxResidueAnalog} can be considered as the sum of a
fast-fading signal and thermal noise, where the fade is given by
$j(\phi_{\sf si}(t -\Delta_{\sf si}) - \phi_{\sf si}(t))$. Since the
fade, $j(\phi_{\sf si}(t -\Delta_{\sf si}) - \phi_{\sf si}(t))$, is zero
mean and changes every sample, it cannot be estimated and thus digital
cancellation cannot reduce the residual self-interference any
further. More precisely,
\begin{equation}
   \mathbb{E}(y_{\sf residue-analog}[iT]x_{\sf si}[iT]) 
 \approx
 h_{\sf si}R_{x_{\sf si}}(\Delta_{\sf si})\mathbb{E}\left(j\phi_{\sf si}[iT] - j\phi_{\sf cancel}[iT - \Delta_{\sf si}]\right)
 \mathbb{E}\left(e^{-j\phi_{\sf down}[iT]}\right) \stackrel{\text{(a)}}{=}  0
 \label{eq:digitaldoesnothelp3}
 \end{equation}
 where (a) is true because phase noise is assumed to be zero
 mean. From \eqref{eq:digitaldoesnothelp3}, it is clear that the
 residual self-interference after active analog cancellation is
 uncorrelated to the self-interference signal and thus digital
 cancellation does not cancel self-interference any further.

The result that the residual self-interference after active analog
cancellation \emph{not} correlated to $x_{\sf si}[iT]$ when perfect
channel estimates are available is not limited to pre-mixer
cancellers. In post-mixer cancellers, perfect estimates for active
analog cancellation imply that the residual is only thermal noise,
which is naturally uncorrelated to the self-interference. In baseband
cancellers, the correlation of the residual and self-interference
signal can be written as
\begin{eqnarray}
& &  \mathbb{E}(y_{\sf residue-analog}[iT]x_{\sf si}[iT]) \nonumber \\
& = & h_{\sf si}\mathbb{E}\left(x_{\sf
  si}[iT]x_{\sf si}[iT - \Delta_{\sf si}](e^{j\phi_{\sf si}[iT]} -
e^{j\phi_{\sf down}[iT - \Delta_{\sf si}]})\right)e^{-j\omega_c\Delta_{\sf si}}
+ \mathbb{E}(x_{\sf si}[iT]z_{\sf noise}[iT])\nonumber \stackrel{\text{(a)}}{=}  0,
\end{eqnarray}
where (a) holds when $\phi_{\sf si}(t)$ and $\phi_{\sf down}(t)$ are
identically distributed. If $\phi_{\sf si}(t)$ and $\phi_{\sf
  down}(t)$ are not distributed identically, then correlation of the
self-interference signal with the residual self-interference is
approximately zero if $\phi_{\sf si}[iT] << 1, \phi_{\sf down}[iT] <<
1$. Digital cancellation is form of active cancellation, much like
active analog cancellation. When perfect channel estimates are
available, successively performing active cancellation is equivalent
to actively cancelling in analog domain once.

\subsection{Digital cancellation when active analog cancellation uses imperfect channel estimate}
%% In Section~\ref{sec:AllCancelersImperfect}, we show that if active
%% analog cancellation is performed with imperfect channel estimates,
%% then the residual self-interference signal has a term which is
%% dependent entirely on the self-interference signal $x_{\sf
%%   si}(iT)$. Here, we show that using imperfect channel estimates for
%% active analog cancellation will lead to a residual which is correlated
%% to $x_{\sf si}(t)$, which can be subtracted by digital cancellation,
%% thereby reducing the strength of the residual.
\noindent \emph{Result 4: If active analog cancellation uses imperfect
  channel estimates, then digital cancellation following it can cancel
  the residual correlated with the self-interference signal, thereby
  reducing its strength. However, the sum of the cascaded stages of
  active cancellation is limited by the phase noise properties and the
  error in channel estimate used for active analog cancellation}

\subsubsection{Pre-mixer canceller}
As an example, let us consider the residual self-interference in
pre-mixer canceller. Let us define the residual self-interference
channel as
\begin{equation} 
\mathbf{h}_{\sf residual-si}[iT] = h_{\sf si}(\delta[iT -
\Delta_{\sf si}]e^{-j\omega_c\Delta_{\sf si}} - \rho \delta[iT
-\tau]e^{-j\omega_c\tau}).
\label{eq:ResidualSIFilter}
\end{equation}
Then, the residual self-interference signal in the digital domain can be
 written as
\begin{eqnarray}
&& \mathbf{h}_{\sf residual-si}[iT]\ast x_{\sf si}[iT] e^{j(\phi_{\sf
      cancel}[iT] - \phi_{\sf down}[iT])} + r_{\sf phase-noise,pre}[iT] + z_{\sf
    noise}[iT],
\label{eq:CompactDigiResidualPre}
\end{eqnarray}
where 
\begin{equation}
  r_{\sf phase-noise,pre}[iT] = jh_{\sf si}e^{-j\omega_c\Delta_{\sf si}}x_{\sf si}[iT - \Delta_{\sf si}](\phi_{\sf si}[iT - \Delta_{\sf si}] - \phi_{\sf
    cancel}[iT])e^{j(\phi_{\sf  cancel}[iT] - \phi_{\sf down}[iT])}
\end{equation}
is the residual which is dependent on phase noise and uncorrelated
with the self-interference signal $x_{\sf si}[iT]$. The digital
canceller can use an estimate of the residual self-interference
channel, $\widehat{\mathbf{h}}_{\sf residual-si}[iT]$, to generate a
cancelling signal,$-\widehat{\mathbf{h}}_{\sf residual-si}[iT]\ast
x_{\sf si}[iT]$,  which will result in a residual
self-interference given by
\begin{eqnarray}
y_{\sf residue-digital}[iT] & = & (\mathbf{h}_{\sf residual-si}[iT])\ast x_{\sf si}[iT]) e^{j(\phi_{\sf
      cancel}[iT] - \phi_{\sf down}[iT])} -  \widehat{\mathbf{h}}_{\sf residual-si}[iT] \ast x_{\sf si}[iT] \nonumber \\ && + r_{\sf phase-noise,pre}[iT] + z_{\sf
    noise}[iT] \nonumber \\
& \approx &(\mathbf{h}_{\sf residual-si}[iT] -  \widehat{\mathbf{h}}_{\sf residual-si}[iT])\ast x_{\sf si}[iT]  + r_{\sf phase-noise,pre}[iT]\nonumber \\
&&        + j \mathbf{h}_{\sf residual-si}[iT] \ast x_{\sf si}[iT](\phi_{\sf cancel}[iT] - \phi_{\sf down}[iT])  + z_{\sf
    noise}[iT].
\label{eq:ImperfectEverythingPre-radioResidue}
\end{eqnarray}
The strength of the residual self-interference after digital
cancellation is
\begin{eqnarray}
&&     \mathbb{E}(|y_{\sf residue-digital}[iT]|^2) \nonumber \\ & \approx & \mathbb{E}(|(\mathbf{h}_{\sf residual-si}[iT] - \widehat{\mathbf{h}}_{\sf residual-si}[iT])\ast x_{\sf si }[iT]|^2) + \mathbb{E}(|r_{\sf phase-noise}[iT]|^2)   \nonumber \\ & + & \mathbb{E}(|(\mathbf{h}_{\sf residual-si}[iT]\ast x_{\sf si }[iT])(\phi_{\sf cancel}[iT] -\phi_{\sf down}[iT])|^2)  + \mathbb{E}(|z_{\sf noise}[iT]|^2) \nonumber \\
  & = &  \underbrace{\mathbb{E}(|(\mathbf{h}_{\sf residual-si}[iT] - \widehat{\mathbf{h}}_{\sf residual-si}[iT])\ast x_{\sf si }[iT]|^2)}_{\text{imperfect estimate in digital domain}}  +   \underbrace{2|h_{\sf si}|^2\sigma_{\sf si}^2(1 - R_{\phi_{\sf si}}(\Delta_{\sf si}))}_{\text{phase noise}} + \sigma_{\sf noise}^2  \nonumber \\ & & + \underbrace{\mathbb{E}(|(\mathbf{h}_{\sf residual-si}[iT]\ast x_{\sf si }[iT])|^2)}_{\text{imperfect estimate in analog domain}}(\sigma_{\sf si}^2 +  \sigma_{\sf down}^2). 
\label{eq:ImperfectEverythingPre-radio}
\end{eqnarray}
We make the following two observations from~\eqref{eq:ImperfectEverythingPre-radio}.

\begin{figure}[htb]
\begin{minipage}[b]{0.45\linewidth}
\centering
  \scalebox{0.45}{\includegraphics[trim= 20mm 65mm 30mm 70mm, clip]{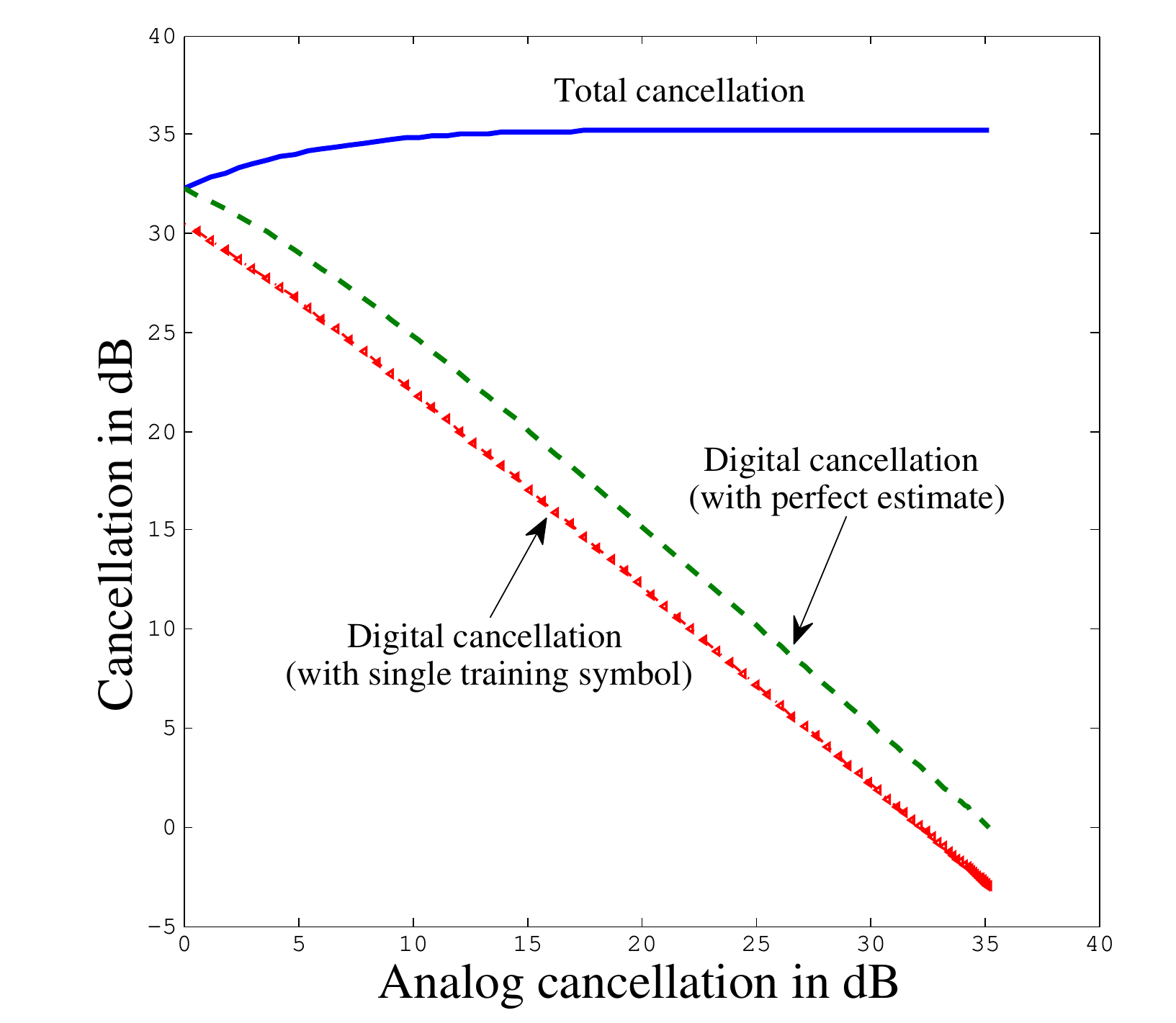}}
\caption{The relationship between amount of active analog cancellation
  and the amount of digital cancellation in a pre-mixer canceller is
  shown. Also, we assume $\sigma_{\sf si}^2 = \sigma_{\sf down}^2$.}
\label{fig:avd}
\end{minipage}
\hspace{0.5cm}
\begin{minipage}[b]{0.45\linewidth}
\centering
\scalebox{0.45}{\includegraphics{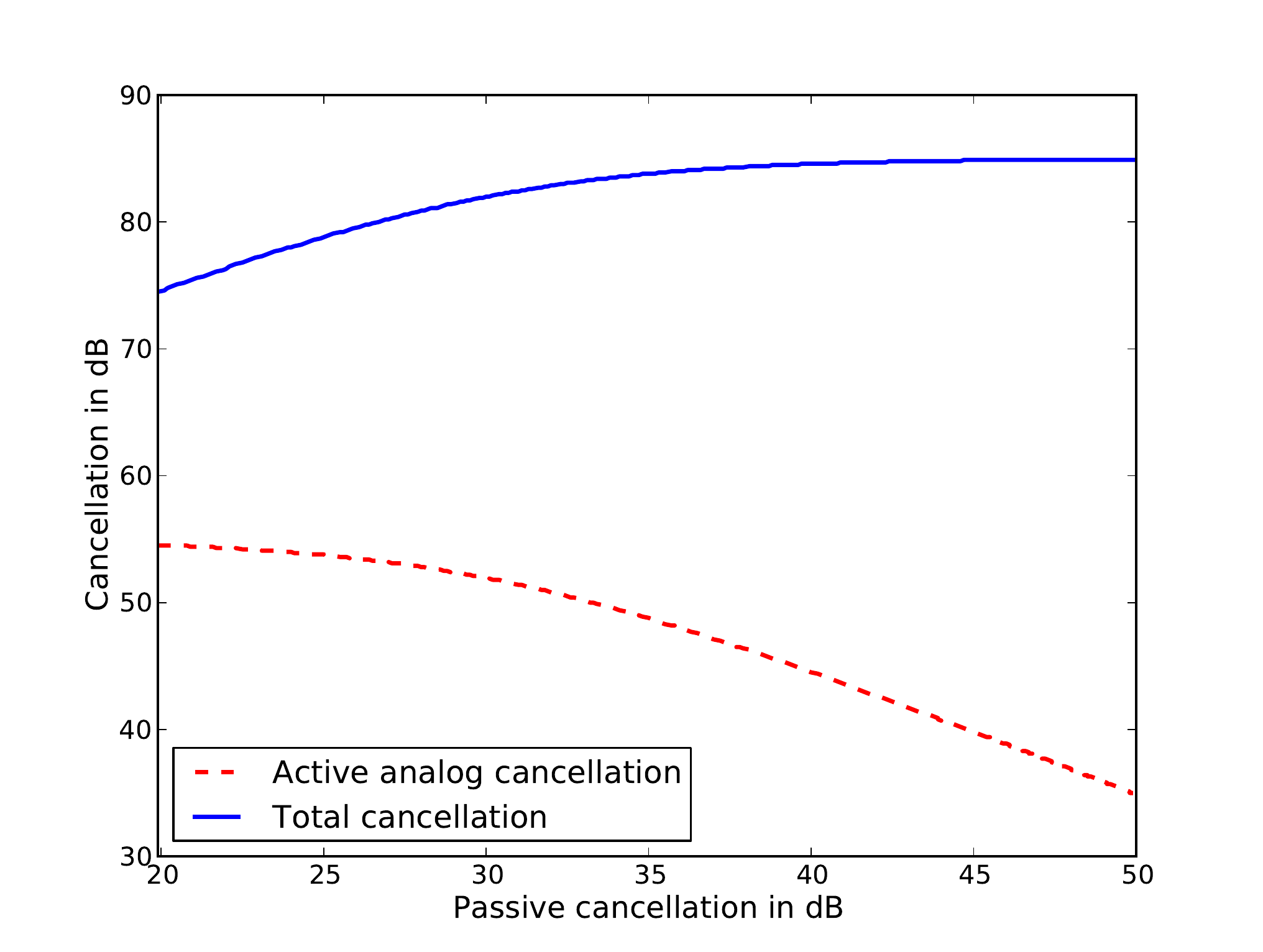}}
 \caption{Total cancellation represents the sum of passive and active
    analog cancellation when operated in cascade in a pre-mixer
    canceller.}
\label{fig:pva}
\end{minipage}
\end{figure}

\emph{Observation 6:} The amount of residual self-interference after
digital cancellation stage is lower bounded by $2|h_{\sf
  si}|^2\sigma_{\sf si}^2(1 - R_{\sf si}(\Delta_{\sf si})) +
\sigma_{\sf noise}^2$, which, we recall from
Section~\ref{sec:Pre-radioPerfect}, is the strength of residual
self-interference after active analog cancellation that uses perfect
estimate of self-interference channel. If the digital canceller uses
perfect estimate of the residual self-interference channel,
$\widehat{\mathbf{h}}_{\sf residual-si}[iT] = \mathbf{h}_{\sf
  residual-si}[iT]$, then it can eliminate the residual that depends
only on self-interference signal entirely.  Figure~\ref{fig:avd} shows
the amount of digital cancellation possible as a function of active
analog cancellation for a pre-mixer canceller where the local
oscillators in the cancelling and self-interference path are
independent which implies that $R_{\phi_{\sf si}}(\Delta_{\sf si}) =
0$. Figure~\ref{fig:avd} explains the trend of active analog
vs.~digital cancellation reported in \cite{Duarte:phd}, that the sum
total active cancellation of active analog and digital stages is no
more than 35~dB, which is the amount of cancellation achieved when the
analog stage uses perfect estimates.
  
%% \item The lower bound in \eqref{eq:DigitalResidue} i.e., when $\rho =
%%   1$ is set in \eqref{eq:DigitalResidue}, tells us that strength of
%%   the residual signal with cascaded stages of digital (with perfect
%%   estimates) and imperfect analog cancellation only as good as analog
%%   cancellation with perfect estimates.
\emph{Observation 7:} If $\sigma_{\sf down}^2 >> \sigma_{\sf si}^2$,
then the receiver phase noise will be a dominant source of bottleneck
in digital cancellation. In computing the contribution of receiver
phase noise to residual self-interference signal, we note that the
variance of receiver phase noise is scaled by strength of the residual
self-interference channel. Poor active analog cancellation implies
that $\mathbb{E}(|(\mathbf{h}_{\sf residual-si}[iT]\ast x_{\sf si
}[iT]|^2)$ is large. Therefore, as depicted in Figure~\ref{fig:avd},
poor active analog cancellation results in less overall cancellation,
even when digital cancellation uses perfect estimate of
self-interference channel.

%% \begin{figure}[!h]
%% \centering
%%   \scalebox{0.6}{\includegraphics[trim= 20mm 60mm 30mm 70mm, clip]{figs/avd.pdf}}
%% \caption{The relationship between amount of active analog cancellation
%%   and the amount of digital cancellation in a pre-radio canceller is
%%   shown. The variance of transmitter and receiver phase noise is
%%   assumed to be same i.e., $\sigma_{\sf si}^2 = \sigma_{\sf down}^2$.}
%% \label{fig:avd}
%% \end{figure}

\subsubsection{Post-mixer cancellers}
In post-mixer cancellers too, the digital cancellation when cascaded
with active analog cancellation can only cancel the portion of
residual self-interference that is correlated with the
self-interference signal itself.

For post-mixer cancellers, the residual self-interference channel is
defined as in \eqref{eq:ResidualSIFilter} and the phase noise
dependent residual self-interference is given by
\begin{equation}
  r_{\sf phase-noise,post} = jh_{\sf si}e^{-j\omega_c\Delta_{\sf si}}x_{\sf si}[iT - \Delta_{\sf si}](\phi_{\sf si}[iT - \Delta_{\sf si}] - \phi_{\sf
    si}[iT - \tau])e^{j(\phi_{\sf  si}[iT -\tau] - \phi_{\sf down}[iT])}.
\end{equation}
The residual self-interference before digital cancellation will be 
\begin{equation}
  \mathbf{h}_{\sf residual-si}[iT]\ast x_{\sf si}[iT] e^{j(\phi_{\sf
      si}[iT -\tau] - \phi_{\sf down}[iT])} + r_{\sf
    phase-noise,post}[iT] + z_{\sf noise}[iT].
\label{eq:CompactDigiResidualPost}
\end{equation}
Note that the form of \eqref{eq:CompactDigiResidualPost} is very
similar to \eqref{eq:CompactDigiResidualPre} and thus, without
repeating the steps, we can write the residual in post-mixer
cancellers after digital cancellation with imperfect estimates as
\begin{eqnarray}
  && \mathbb{E}(|y_{\sf residual-digital}[iT]|^2) \nonumber \\
  & = & \mathbb{E}(|(\mathbf{h}_{\sf residual-si}[iT] - \widehat{\mathbf{h}}_{\sf residual-si}[iT])\ast x_{\sf si }[iT]|^2) +   2|h_{\sf si}|^2\sigma_{\sf si}^2(1 - R_{\phi_{\sf si}}(\Delta_{\sf si} - \tau)) + \sigma_{\sf noise}^2  \nonumber \\ & & + \mathbb{E}(|(\mathbf{h}_{\sf residual-si}[iT]\ast x_{\sf si }[iT])|^2)(\sigma_{\sf si}^2 +  \sigma_{\sf down}^2)
\label{eq:ImperfectDigitalPost}
\end{eqnarray}
Note that \eqref{eq:ImperfectDigitalPost} is lower bounded by
$2|h_{\sf si}|^2\sigma_{\sf si}^2(1 - R_{\phi_{\sf si}}(\Delta_{\sf
  si} - \tau)) + \sigma_{\sf noise}^2$, which itself is the lower
bound on the strength of the residual when active analog cancellation
uses imperfect estimate of the channel. Thus, even in post-mixer
cancellers, more digital cancellation is possible when active analog
cancellation cancels less. However, the sum of cancellation is no more
than $1/(2\sigma_{\sf si}^2(1 - R_{\phi_{\sf si}}(\Delta_{\sf si} -
\tau))$, an expression which is solely dependent on phase noise.

\subsubsection{Baseband analog cancellers}
For baseband analog cancellers, let the residual self-interference
channel be defined as in \eqref{eq:ResidualSIFilter}, and the
residual dependent on phase noise be given by
\begin{equation}
  r_{\sf phase-noise,bb}[iT] = jh_{\sf
    si}e^{-j\omega_c\Delta_{\sf si}}x_{\sf si}[iT - \Delta_{\sf
    si}](\phi_{\sf si}[iT - \Delta_{\sf si}] - \phi_{\sf down}[iT]).
\end{equation}
The residual self-interference before digital cancellation be written
as
\begin{equation}
  \mathbf{h}_{\sf residual-si}[iT]\ast x_{\sf si}[iT] + r_{\sf phase-noise,bb}[iT] + z_{\sf noise}[iT].
\end{equation}
Using $\widehat{\mathbf{h}}_{\sf residual-si}[iT]\ast x_{\sf si}[iT]$
as the cancelling signal, the strength of residual after imperfect
digital cancellation is given by
\begin{eqnarray}
&& \mathbb{E}(|y_{\sf residue-digital}[iT]|^2)\nonumber \\ & = &
  \mathbb{E}(|(\mathbf{h}_{\sf residual-si}[iT] -
  \widehat{\mathbf{h}}_{\sf residual-si}[iT])\ast x_{\sf si}[iT]|^2) +
  \mathbb{E}(|r_{\sf phase-noise,bb}[iT]|^2) + \mathbb{E}(|z_{\sf
    noise}[iT]|^2) \nonumber \\ & = & \mathbb{E}(|(\mathbf{h}_{\sf
    residual-si}[iT] - \widehat{\mathbf{h}}_{\sf residual-si}[iT])\ast
  x_{\sf si}[iT]|^2) + |h_{\sf si}|^2(\sigma_{\sf si}^2 + \sigma_{\sf
    down}^2) + \sigma_{\sf noise}^2 \nonumber \\
& \geq & |h_{\sf si}|^2(\sigma_{\sf si}^2 + \sigma_{\sf
    down}^2) + \sigma_{\sf noise}^2.
\label{eq:DigitalResidueBB}
\end{eqnarray}
The lower bound in \eqref{eq:DigitalResidueBB} is the strength of
residual self-interference after active analog cancellation is
performed with perfect channel estimates in baseband cancellers. The
lower bound in \eqref{eq:DigitalResidueBB} is achievable if the
digital canceller has perfect estimate of the residual
self-interference channel. Thus, serially concatenated active analog
cancellation and digital cancellation are interdependent in such way
that their sum is bounded by $\frac{1}{\sigma_{\sf si}^2 + \sigma_{\sf
    down}^2}$. One distinction in baseband analog cancellers is that
unlike pre-mixer or post-mixer cancellers, the residual does not
depend explicitly on the quality of active analog cancellation, i.e.,
$\mathbf{h}_{\sf residual-si}[iT]$, rather is dependent upon
$(\mathbf{h}_{\sf residual-si}[iT] - \widehat{\mathbf{h}}_{\sf
  residual-si}[iT])$, the quality of digital cancellation only.

\section{Answer 3. Influence of Passive Suppression on Active Cancellation}
\label{sec:PassiveVsActive}
In this section, we answer ``How and when does passive suppression
impact the amount of analog cancellation?'' We show that the amount of
passive suppression can impact the amount of active analog
cancellation in pre-mixer cancellers.

So far, we have considered a self-interference channel with only a
single delay tap. Now, let us consider a self-interference channel
with two non-zero taps, which can be considered as taps representing
line of sight and the reflected components. Let the two-tap
self-interference channel be $\mathbf{h}_{\sf si}(t) = h_1\delta(t -
\Delta_1) + h_2\delta(t - \Delta_2)$, where $\Delta_1$ and $\Delta_2$
denote the delays of the line of sight and reflected component,
therefore $\Delta_{1} < \Delta_{2}$. The average strength of the line
of sight and reflected component are captured by $\mathbb{E}(|h_1|^2)$
and $\mathbb{E}(|h_2|^2)$. It is reasonable to assume that passive
suppression can reduce the strength of the line of sight
component. Therefore, the amount of passive suppression determines the
ratio $\mathbb{E}(|h_1|^2)/\mathbb{E}(|h_2|^2)$.

%% As an example, if line of sight path is significantly suppressed
%% then $\mathbb{E}(|h_1|^2) << \mathbb{E}(|h_2|^2)$. {\bf Achal, this
%% is a little awkward statement, only the ratio of h1 and h2 will
%% significantly change, but saying something about absolute
%% magnitudes is misleading.}
\noindent \emph{Result 5: Higher passive suppression can result in
  lower active analog cancellation in pre-mixer cancellers. However,
  increasing passive suppression implies that sum of cascaded passive
  and active analog cancellation increases.}

Assume self-interference channel is perfectly known. Then the
cancelling signal in baseband is 
\begin{equation}
x_{\sf cancel}(t) = -h_{1}x_{\sf si}(t -
\Delta_{1})e^{-j\omega_c\Delta_{1}} -
h_2x_{\sf si}(t -\Delta_{2})e^{-j\omega_c\Delta_{2}}.
\end{equation} 
In presence of phase noise, the residual self-interference is
\begin{eqnarray}
  && y_{\sf residual}(t) \nonumber \\ &  = &  h_{1}x_{\sf si}(t -
  \Delta_{1})e^{-j\omega_c\Delta_{1}}(e^{j\phi(t - \Delta_1)} -
  e^{j\phi(t)})  +  h_2x_{\sf si}(t  -\Delta_{2})e^{-j\omega_c\Delta_{2}}(e^{j\phi(t - \Delta_2)} - e^{j\phi(t)}) +z_{\sf noise}(t) \nonumber \\
& \approx & jh_{1}x_{\sf si}(t -
  \Delta_{1})e^{-j\omega_c\Delta_{1}}(\phi(t - \Delta_1) - \phi(t)) + jh_{2}x_{\sf si}(t -
  \Delta_{2})e^{-j\omega_c\Delta_{2}}(\phi(t - \Delta_2) - \phi(t))  + z_{\sf noise}(t). \nonumber 
\end{eqnarray}
The strength of the residual signal is 
\begin{eqnarray}
  \mathbb{E}(|y_{\sf residual}(t)|^2) & \approx &  2\mathbb{E}(|h_1|^2)(1 -
  R_{\phi_{\sf si}}(\Delta_1)) + 2\mathbb{E}(|h_2|^2)(1 - R_{\phi_{\sf
      si}}(\Delta_2)) \nonumber \\
 & &  +
  2\mathbb{E}\left(\text{Re}(h_1h_2'x_{\sf si}(t - \Delta_1)x'_{\sf si}(t -
  \Delta_2)e^{j\omega_c(\Delta_2 - \Delta_1)})\right) \nonumber \\
 && (1 + R_{\phi_{\sf
      si}}(\Delta_1 - \Delta_2) - R_{\phi_{\sf si}}(\Delta_1) - R_{\phi_{\sf si}}(\Delta_2))\sigma_{\phi}^2.
\end{eqnarray}
The average residual self-interference can be estimated by assuming a
distribution on the line of sight and the reflected component
channel. From the experimental characterization of the
self-interference channel in \cite{Duarte:2011aa}, we know that when
the line of sight component is sufficiently suppressed, the
self-interference channel is approximately a zero mean complex
Gaussian random variable. Therefore, we have
\begin{equation}
  \mathbb{E}(|y_{\mathsf{residual}}(t)|^2) = \mathbb{E}(|h_1|^2)(1 -
  R_{\phi_{\mathsf{si}}}(\Delta_1)) + \mathbb{E}(|h_2|^2)(1 -
  R_{\phi_{\mathsf{si}}}(\Delta_2)),
\label{eq:ResidueTwoTap}
\end{equation}
assuming the independence of $h_1$ and $h_2$. If either
$\mathbb{E}(|h_1|^2)$ or $\mathbb{E}(|h_2|^2)$ is reduced, it amounts
to increasing the passive suppression. The design principle that
increasing passive suppression reduces total residual
self-interference is confirmed by equation \eqref{eq:ResidueTwoTap}
and it is also depicted in Figure~\ref{fig:pva}.

The amount of active analog cancellation is obtained by computing the
ratio of the strength of self-interference before and after active
analog cancellation which is
\begin{equation}
  \frac{\mathbb{E}(|h_1|^2 + |h_2|^2)}{\mathbb{E}(|h_1|^2)(1  - R_{\phi_{\sf si}}(\Delta_1)) + \mathbb{E}(|h_2|^2)(1 - R_{\phi_{\sf si}}(\Delta_2))}.
\end{equation}
The strength of the line of sight component, $\mathbb{E}(|h_1|^2)$,
varies as the coupling between transmit and receive antenna on the
full-duplex node changes. At one extreme if passive suppression is low
and line of sight is dominant,
$\mathbb{E}(|h_1|^2)/\mathbb{E}(|h_2|^2) >> 1$, then the amount of
active analog cancellation possible is $1/(1 - R_{\phi_{\sf
    si}}(\Delta_1))$. At the other extreme, if passive suppression is
very high and the strength of line of sight component is negligible,
then the amount of active analog cancellation possible is $1/(1 -
R_{\phi_{\sf si}}(\Delta_2))$. Thus, amount of passive suppression
influences the amount of active analog cancellation. Moreover, since
$\Delta_1 < \Delta_2$ implies $1/(1 - R_{\phi_{\sf si}}(\Delta_1)) >
1/(1 - R_{\phi_{\sf si}}(\Delta_2))$, thus more passive suppression
implies less active analog cancellation. In Figure~\ref{fig:pva} we
plot the amount of active cancellation as a function of the strength
of the line of sight component. Note that the total cancellation is
maximized when passive suppression is maximum, however active analog
cancellation reduces as passive suppression increases.

  %% {\bf Achal, from the above
  %% discussion, it was not clear how phase noise played a role here. I
  %% know it is part of the analysis. Also, what about other cancelers?
  %% Do you want to say that the analysis and results will extend to
  %% those too due to similarities noted in previous sections?}
%% \begin{figure}[!h]
%%   \centering
%%   \scalebox{0.5}{\includegraphics{figs/passivevsactive.pdf}}
%%   \caption{Total cancellation represents the sum of passive and active
%%     analog cancellation when operated in cascade in a pre-radio
%%     canceller.}
%% \label{fig:pva}
%% \end{figure}

\section{Signal Model for Full-Duplex}
\label{sec:Model}
Using the analyses in Sections \ref{sec:AllCancellers} and
\ref{sec:DigitalCancellation}, we develop a signal model for SISO
full-duplex communication, and then extend it to the MIMO and wideband
cases.
\subsection{Narrowband signal model}
\label{sec:model-narrowband}
We present a digital baseband signal model which captures the effect
of phase noise and imperfection in channel estimate by considering the
residual self-interference after: (a) active analog cancellation and
(b) digital cancellation cascaded with active analog cancellation. For
both (a) and (b), the channel estimates are assumed to be imperfect.
%% The signal model due to (a) captures the performance of full-duplex
%% systems which have no digital cancellation, while (b) is indicative of
%% performance of full-duplex systems with both active analog and digital
%% cancellations.
\paragraph{Active analog cancellation with imperfect estimates}
For pre-mixer cancellers the residual self-interference is given by
\eqref{eq:ApproxImperfectAnalogResidue}. Since phase noise is assumed
to be zero mean Gaussian, the linear combination of several phase
noise terms is also Gaussian. Also, phase noise is assumed to be
small, therefore $e^{j(\phi_{\sf si}(t - \Delta_{\sf si}) - \phi_{\sf
    cancel}(t))} \approx 1$. Then the received signal at $\sf N$1,
which is a combination of residual self-interference, signal of
interest and thermal can be written as
\begin{eqnarray}
  y_{1}[iT] &  = & \sqrt{P_{\sf signal}}\mathbf{h}_{\sf signal}[iT]\ast x_{\sf
    signal}[iT] + \sqrt{P_{\sf si}}|h_{\sf si}|\beta_{\phi}z_{\sf
    phase-noise}[iT] \nonumber\\ & & + \sqrt{P_{\sf si}}\mathbf{h}_{\sf residual-si}[iT]\ast x_{\sf si}[iT] + z_{\sf noise}[iT],
\label{eq:DiscreteTimeModel1}
\end{eqnarray}
where $z_{\sf phase-noise}[iT]$ is a zero mean AWGN with unit variance
independent of the thermal noise and signal of interest. The signal
$x_{\sf signal}[iT]$ is of unit variance and $P_{\sf si}$ and $P_{\sf
  signal}$ are power constraints at $\sf N$1 and $\sf N$2
respectively. The contribution of phase noise to the residual
self-interference is captured by $\beta_{\phi}$ whose value is given
in Table~\ref{table:model}.

For post-mixer cancellers, as well as baseband analog cancellers, the
contribution of phase noise to the residual is different than
pre-mixer cancellers. However, the form of the residual
self-interference after active analog cancellation in post-mixer and
baseband analog cancellers is given by
\eqref{eq:PostRadioAnalogResidue} and
\eqref{eq:ImperfectBasebandResidue} respectively, which is similar to
\eqref{eq:ApproxImperfectAnalogResidue}. Therefore the signal model
\eqref{eq:DiscreteTimeModel1} holds for post-mixer and baseband analog
cancellers too. The parameter $\beta_{\phi}$ for each canceller can be
obtained from Table \ref{table:model}.

\paragraph{Imperfect estimates in active analog and digital cancellation}
After digital cancellation, the residual depends on the quality of the
estimate of residual self-interference channel, in addition to phase
noise. For pre-mixer cancellers, the residual is given by
\eqref{eq:ImperfectEverythingPre-radioResidue} and the strength of the
residual is given by \eqref{eq:ImperfectEverythingPre-radio}, which
allows us to write the received signal at $\sf N$1 as
\begin{eqnarray}
  y_{1}[iT] & = & \sqrt{P_{\sf signal}}\mathbf{h}_{\sf signal}[iT]\ast x_{\sf signal}[iT] + \sqrt{P_{\sf si}}|h_{\sf si}|\gamma_{\phi}z_{\sf
    phase-noise}[iT]  \nonumber \\ && + \sqrt{P_{\sf si}}(\mathbf{h}_{\sf residual-si}[iT] - \widehat{\mathbf{h}}_{\sf residual-si}[iT])\ast x_{\sf si}[iT])  + z_{\sf noise}[iT],
  \label{eq:DiscreteTimeModel2}
\end{eqnarray}
where $\gamma_{\phi}$ is a parameter dependent on the phase noise and
the quality of active analog cancellation. For post-mixer and baseband
analog cancellers, the signal model in \eqref{eq:DiscreteTimeModel2}
is modified appropriately by changing the parameter $\gamma_{\phi}$,
which is computed in \eqref{eq:ImperfectDigitalPost} and
\eqref{eq:DigitalResidueBB} respectively, and populated in
Table~\ref{table:model}.

\subsection{Wideband signal model}
\label{sec:model-wideband}
Wideband full-duplex is implemented in \cite{achal:fd,Choi:2011m}. In
wideband full-duplex, the self-interference channel need not be
frequency flat \cite{Duarte:2011aa,evan}. To derive a signal model for
wideband full-duplex, we treat it as a combination of several
narrowband full-duplex systems. Let the overall bandwidth be $W$ and
the coherence bandwidth of the self-interference channel be $w$, then
wideband channel is composed of $K = \lceil \frac{W}{w} \rceil$
narrowband channels. Let the $k^{\text{th}}$ narrowband channel be
denoted by $h_{{\sf si},k}\delta(t - \Delta_{{\sf si},k})$. If the
bandwidth is much smaller than the carrier frequency, $W << \omega_c$,
then the phase jitter over the band of interest can be assumed to be
independent of the bandwidth \cite{ad:phasenoise}.

The signal model for wideband full-duplex can be described by
explicitly writing the expression for received signal in each of $K$
narrowband channels. After active analog cancellation with imperfect
channel estimate, the received signal in the $k^{\rm th}$ narrrowband
channel is given by
\begin{eqnarray}
  y_{1,k}[iT]& = & \mathbf{h}_{\mathsf{signal}, k}[iT]\ast
  x_{\mathsf{signal},k}[iT] +
  \sqrt{P_{\mathsf{si},k}}|h_{\mathsf{si},k}|\gamma_{\phi}z_{\mathsf{phase-noise},k}[iT]
  \nonumber \\ && + \mathbf{h}_{\mathsf{residual-si},k}[iT]\ast x_{\mathsf{si},k}[iT] + z_{\sf noise}[iT],
  \label{eq:DiscreteTimeModel3}
\end{eqnarray}
where $P_{\mathsf{si},k}$ is power constraint for each band. Note
that, while the phase noise in each band scales according to the
transmit power in that band, the thermal noise floor remains constant.
To compare the bottleneck in narrowband vs.~wideband let us assume the
total power in both is the same, say $P$. As a simplifying assumption,
let $|h_{\sf si,k}| = |h_{\sf si}|$. In the narrowband system, the
strength of residual self-interference due to phase noise is $P|h_{\sf
  si}|^2\beta_{\phi}^2$, which is the same as the strength of the
residual self-interference due to phase noise is in wideband, i.e.,
$\sum_{i = 1}^K P_{\sf si,k}|h_{\sf si,k}|^2\beta_{\phi}^2 = P|h_{\sf
  si}|^2\gamma_{\phi}^2$. On the other hand, if the thermal noise
floor in narrowband is given by the variance $\sigma_{\sf noise}^2$,
then the variance of the noise over wideband is $K\sigma_{\sf
  noise}^2$. The signal model after digital cancellation can be
written by simply replacing $\beta_{\phi}$ by $\gamma_{\phi}$, and $
\mathbf{h}_{\mathsf{residual-si},k}[iT]$ with $\widehat{
  \mathbf{h}}_{\mathsf{residual-si},k}[iT]$ in
\eqref{eq:DiscreteTimeModel3}.

\subsection{MIMO full-duplex signal model}
\label{sec:model-mimo}
To extend the narrowband SISO model \eqref{eq:DiscreteTimeModel1}, we
assume a MIMO system with $M$ transmit antenna and $N$ received
antenna. The self-interference at each of the receivers is due to the
sum of $M$ transmissions, one from each transmit antenna. If the
transmit radio chain for each antenna has an independent local
oscillator, then the residual self-interference due to phase noise is
the sum of $M$ independent residuals due to phase noise in a SISO
system. Then, the received signal at the $n^{\rm th}$ receiver of the
full-duplex node $\sf N$1 is given as
\begin{eqnarray}
  y_{1,n}[iT]& = &\sum_{m = 1}^M \sqrt{P_{\mathsf{signal},m}}\mathbf{h}_{\mathsf{signal}, mn}[iT] \ast
  x_{\mathsf{signal}, m}[iT] + \gamma_{\phi}\sqrt{\sum_{m = 1}^M|h_{\mathsf{si},m}|^2P_{\mathsf{si},m}}z_{\mathsf{phase-noise},n}[iT] \nonumber \\ && +
  \sum_{i = 1}^M \mathbf{h}_{\mathsf{residual-si},mn}[iT]\ast x_{\sf si,m}[iT] +
  z_{\mathsf{noise},n}[iT],
\label{eq:MIMOmodel1}
\end{eqnarray}
where $z_{\mathsf{phase-noise},n}[iT]$ is unit variance, while
$z_{\mathsf{noise},n}[iT]$ has a variance of $\sigma_{\sf
  noise}^2$. The $\mathbf{h}_{\mathsf{signal},mn}[iT]$ represents the
channel for the signal of interest from $m^{\rm th}$ transmitter to
$n^{\rm th}$ receiver. The self-interference channel and the residual
self-interference channel at $\sf N$1 is represented by
$\mathbf{h}_{\mathsf{si},mn}[iT]$ and
$\mathbf{h}_{\mathsf{residual-si},mn}[iT]$ respectively. Power
constraints at the $m^{\rm th}$ transmitter for the signal of interest
and self-interference is $P_{\mathsf{signal},m}$ and
$P_{\mathsf{si},m}$ respectively. To qualitatively understand the MIMO
model in \eqref{eq:MIMOmodel1}, consider the special case where all
the self-interference channels have identical magnitude, the residual
self-interference is simply $M$ times the residual self-interference
for SISO. To describe the signal model after digital cancellation, we
can extend the signal model in \eqref{eq:MIMOmodel1} by 
following the steps used to extend
\eqref{eq:DiscreteTimeModel1} to \eqref{eq:DiscreteTimeModel2}.

%\subsection{Role of the new signal model}
%The signal models proposed in this section abstract away the
%architecture of the canceller being used in a full-duplex system by
%replacing it with a residual self-interference which depends on the
%phase noise, the quality of the self-interference channel estimate and
%the strength of the self-interference channel. The new type of
%interference in full-duplex i.e., residual self-interference which
%scales with the strength of self-interference itself is captured by
%the signal models. The signal models can be used to study optimal
%coding schemes in full-duplex. The MIMO signal model can be employed
%to study the degrees of freedom of full-duplex as the number of
%transmit and receive antennas increase.  In general, the signal models
%can be used to study and design communication strategies, while
%capturing the main sources of noise in full-duplex.

\section{Conclusion\label{sec:conclusions}} 

In this paper, we provided an analytical explanation of experimentally observed performance bottlenecks in full-duplex systems. Our analysis clearly shows that phase noise is a  major bottleneck today and thus reducing the phase noise figure of radio mixers could lead to improved self-interference cancellation. 
\section{Appendix}
\subsection{Lower bound for autocorrelation function}
\label{sec:Lowerbound}
Let $S(f)$ be power spectral density of the bandlimited function
$x(t)$ such that $S(f) = 0$ outside $[-F/2, F/2]$.  Due to the power
constraint, we have $\int\limits _{-F/2}^{F/2} S(f)df = 1$. To
evaluate the autocorrelation function $R(.)$ at $\tau$
\begin{eqnarray}
R(\tau) &  = & \int\limits_{-\infty}^{\infty}S(f)e^{-j2\pi f \tau}df  =  \int\limits_{-F/2}^{F/2}S(f)e^{-j2\pi f \tau}df  =  2\int\limits_{0}^{F/2}S(f)\cos{(2\pi f \tau)}df \nonumber \\ &\stackrel{\text{(a)}}{\approx} & 2\int\limits_{0}^{F/2}S(f)(1 - c_1 f^2\tau^2)df  =  1 - c_1\tau^2 \int\limits_{0}^{F/2}f^2 S(f)df \geq  1 - (c_1 F^2/4)\tau^2\left( 2\int\limits_{0}^{F/2} S(f)df \right)\nonumber \\
& = & 1 - c\tau^2,
\end{eqnarray}
where (a) holds if $\tau$ is small, and $c = c_1 F^2/4$.

\subsection{Estimating the suitable scaling for cancellation for delay $d$}
\label{sec:DelayScaling}
Let us denote by $a_1 = h_1e^{-j(\omega_c + \omega)\Delta_1}$ and $a_2
= h_2e^{-j(\omega_c + \omega)\Delta_2}$. If
\eqref{eq:MimicCancellation1} and \eqref{eq:MimicCancellation2} is
true, then
\begin{eqnarray}
  h_c(d) & = & \frac{\sum_{i = 1}^N y_2[(i-d)T]'y_1[iT]}{\sum_{i = 1}^N |y_2[(i-d)T]|^2} \nonumber \\
 & = & \frac{\sum_{i = 1}^N (a_2'x[(i-d)T]' + z_2[(i-d)T]')(a_1x[iT] + z_1[iT])}{\sum_{i = 1}^N(a_2'x[(i -d)T]' + z_2[(i -d)T]')(a_2x[(i -d)T] + z_2[(i -d)T])} \nonumber \\
 & = & \frac{\sum_{i = 1}^N (a_2'e^{j\omega d T}x[iT]' + z_2[(i-d)T]')(a_1x[iT] + z_1[iT])}{\sum_{i = 1}^N(a_2'x[(i -d)T]' + z_2[(i -d)T]')(a_2x[(i -d)T] + z_2[(i -d)T])} \nonumber \\
& = & \frac{\sum_{i = 1}^N (a_2'a_1e^{j\omega d T}|x[iT]|^2 + a_1x[iT]z_2[(i-d)T]' + a_2e^{-j\omega dT}x[iT]'z_1[iT] + z_2[(i - d)T]'z_1[iT])}{\sum_{i = 1}^N(|a_2|^2|x_2[iT]|^2 + |z_2[iT]|^2 + 2\text{Re}\{a_2x_2[iT]z_2[iT]'\})} \nonumber \\
\end{eqnarray}
Letting $N \to \infty$ we can replace the summations with
expectations. Due to independence of thermal noise and the signal, we have
\begin{eqnarray}
  h_c & = & \frac{a_2'a_1e^{j\omega dT}}{|a_2|^2 + \sigma_{\sf noise}^2}  =  \frac{a_1}{a_2}e^{j \omega d T}\left(\frac{1}{1 + \frac{\sigma_{\sf noise}^2}{|a_2|^2}}\right) \approx  \frac{h_1}{h_2}e^{j ((\omega_c + \omega)(\Delta_2 - \Delta_1) + \omega d T)}\left(1 - \frac{\sigma_{\sf noise}^2}{|h_2|^2}\right)
\end{eqnarray}

\subsection{Calculating variance of phase noise}
\label{sec:VarPhaseNoise}
We derive the jitter from the spectrum of the phase noise as
follows. Let the carrier frequency be denoted by $f_c$ and let the
spectrum of the phase noise be specified as $\mathcal{L}(f)$ dBc/Hz
where f is the frequency offset from the carrier frequency.  The phase
jitter in radians is given by $\Delta \theta_{\sf RMS} =
\sqrt{\int_{f_1}^{f_2} 10^{\frac{\mathcal{L}(f)}{10}}df}$, where $f_2
- f_1$ would be bandwidth of the signal ($f_1$ being the lower offset
and $f_2$ being the higher offset).  Jitter in time is given by
$\Delta t_{\sf RMS} = \frac{\Delta \theta_{\sf RMS}}{2\pi f_c}$ and
the corresponding jitter in phase can be calculated as $ \Delta
\theta_{\sf RMS} = \frac{2 \pi f_c \Delta t_{\sf RMS}}{\pi}$. For WARP
radio, MAXIM 2829 \cite{maxim}, operating at a carrier frequency of
2.4GHz results in a time jitter of 0.83 picoseconds which corresponds
to $\sigma_{\phi}=0.717^\circ$, and for the signal generator
\cite{siggen}, operating at 2.2 GHz the phase noise variance is
computed to be $\sigma_{\phi} = 0.066$.

\subsection{Residual computations after active analog cancellations}
\label{sec:ResidualComputations}
%% We compute the residual self-interference after active analog
%% cancellation which uses imperfect estimate of the self-interference
%% channel for different active analog cancellers.
\noindent \emph{Pre-mixer canceller:}
The residual is given by
\eqref{eq:ResiduePreRadioRep}, which can be written as
\begin{eqnarray}
	y_{\sf residue-analog}(t) & = &  h_{\sf
          si}e^{j(\phi_{\sf cancel}(t) -\phi_{\sf down}(t)) }\left(x_{\sf si}(t -
        \Delta_{\sf si})e^{-j\omega_c \Delta_{\sf si}}e^{j(\phi_{\sf
            si}(t - \Delta_{\sf si}) - \phi_{\sf cancel}(t))} - \rho
        x_{\sf si}(t - \tau)e^{-j\omega_c\tau}\right)  \nonumber \\ && +  z_{\sf noise}(t) \nonumber \\ 
         &  \approx &  h_{\sf si}e^{j(\phi_{\sf cancel}(t) - \phi_{\sf down}(t))} \underbrace{(x_{\sf si}(t -
        \Delta_{\sf si})e^{-j\omega_c\Delta_{\sf si}} - \rho
        x_{\sf si}(t - \tau)e^{-j\omega_c\tau})}_{\text{imperfect estimate} } \nonumber \\ &&  + h_{\sf si}e^{j\phi_{\sf cancel}(t)}e^{-j\omega_c\Delta_{\sf si}}x_{\sf si}(t -\Delta_{\sf si})\underbrace{\left(\phi_{\sf si}(t - \Delta_{\sf si}) - \phi_{\sf cancel}(t)\right)}_{\text{phase noise}}  + z_{\sf noise}(t).
 %% &  = &  \left(\underbrace{x_{\sf si}(t - \Delta_{\sf si})e^{j\omega_c(\tau - \Delta_{\sf si})} - \rho x_{\sf si}(t - \tau)  - j \underbrace{\rho x_{\sf si}(t - \tau)(\phi_{\sf cancel}(t) -  \phi_{\sf si}(t - \Delta_{\sf si}))}_{\text{phase noise}}\right)  \nonumber \\  &&
 %%        h_{\sf si}e^{-j\omega_c\tau}e^{j\phi_{\sf si}(t - \Delta_{\sf si})}  + z_{\sf noise}(t).
\label{eq:ApproxImperfectAnalogResidue}
\end{eqnarray}
The strength of the residual is given by 
\begin{eqnarray}
&&	\mathbb{E}(|y_{\sf residue-analog}(t)|^2) \nonumber \\ 
& \approx & |h_{\sf
          si}|^2 (1 + \rho^2 - 2R_{x_{\sf si}}(\Delta_{\sf
          si} - \tau)\text{Re}\{\rho e^{-j\omega_c(\Delta_{\sf si} -\tau)}\})  +  2 \sigma_{\sf si}^2(1  - R_{\phi_{\sf si}}(\Delta_{\sf si}))
             + \sigma_{\sf noise}^2
        \nonumber \\ 
& \geq & |h_{\sf si}|^2( \underbrace{1 +
          \rho^2 - 2|\rho |R_{x_{\sf si}}(\Delta_{\sf si} -
          \tau)}_{\text{imperfect estimate}} +
        \underbrace{2 \sigma_{\sf si}^2(1  - R_{\phi_{\sf si}}(\Delta_{\sf si}))}_{\text{phase noise}})  + \sigma_{\sf noise}^2.
\label{eq:ImperfectAnalogResidueStrength}
\end{eqnarray}
\noindent\emph{Post-mixer canceller:}
The residual self-interference is given by
\begin{eqnarray}
  y_{\mathsf{residue-analog}}(t) & = & h_{\sf si}(x_{\sf si}(t  - \Delta_{\sf si})e^{-j\omega_c\Delta_{\sf si}}e^{j\phi_{\sf si}(t - \Delta_{\sf si})} - \rho x_{\sf si}(t  - \tau)e^{-j\omega_c\tau}e^{j\phi_{\sf si}(t  - \tau)})e^{-j\phi_{\sf down}(t)} \nonumber \\ &&+ z_{\sf noise}(t) \nonumber
\end{eqnarray}
\begin{eqnarray}
& \approx & h_{\sf si}e^{j(\phi_{\sf si}(t - \tau) - \phi_{\sf down}(t))} \underbrace{(x_{\sf si}(t -
        \Delta_{\sf si})e^{-j\omega_c\Delta_{\sf si}} - \rho
        x_{\sf si}(t - \tau)e^{-j\omega_c\tau})}_{\text{imperfect estimate} } \nonumber \\ &&  + h_{\sf si}e^{j\phi_{\sf si}(t - \tau)}e^{-j\omega_c\Delta_{\sf si}}x_{\sf si}(t -\Delta_{\sf si})\underbrace{\left(\phi_{\sf si}(t - \Delta_{\sf si}) - \phi_{\sf si}(t -\tau)\right)}_{\text{phase noise}}  + z_{\sf noise}(t).
\label{eq:PostRadioAnalogResidue}
\end{eqnarray}
and its strength is given by

\begin{equation}
  \mathbb{E}(|y_{\sf residue-analog}(t)|^2) \geq |h_{\sf si}|^2(\underbrace{1 +
  \rho^2 - 2|\rho| R_{x_{\sf si}}(\Delta_{\sf si} - \tau)}_{\text{imperfect estimate}} +
  \underbrace{2\sigma_{\sf si}^2(1 - R_{\phi_{\sf si}} (\Delta_{\sf si} -\tau))}_{\text{phase noise}}) + \sigma_{\sf noise}^2.
\label{eq:PostRadioImperfect}
\end{equation}
\noindent \emph{Baseband analog canceller:}
In baseband analog canceller, the residual self-interference is given by 
\begin{eqnarray}
  y_{\sf residue-analog}(t) & = & h_{\sf si}e^{-j\omega_c\Delta_{\sf
      si}}x_{\sf si}(t - \Delta_{\sf si})(e^{j(\phi_{\sf si}(t -
    \Delta_{\sf si}) - \phi_{\sf down}(t))}) - \rho h_{\sf
    si}e^{-j\omega_c\tau}x_{\sf si}(t - \tau) + z_{\sf noise}(t)
  \nonumber \\
 & \approx & h_{\sf si}(e^{-j\omega_c\Delta_{\sf
      si}}x_{\sf si}(t - \Delta_{\sf si}) - \rho e^{-j\omega_c\tau}x_{\sf
    si}(t - \tau)) \nonumber \\ &&  +  h_{\sf si}e^{-j\omega_c\Delta_{\sf si}}x_{\sf si}(t - \Delta_{\sf
    si})(j(\phi_{\sf si}(t - \Delta_{\sf si}) - \phi_{\sf down}(t))) +
  z_{\sf noise}(t).
  \label{eq:ImperfectBasebandResidue}
\end{eqnarray}
and it's strength is given by
\begin{equation}
\mathbb{E}(|y_{\sf residue-analog}(t)|^2) = |h_{\sf si}|^2
(\underbrace{1 + \rho^2 - 2|\rho| R_{x_{\sf si}}(\tau - \Delta_{\sf
    si})}_{\text{imperfect estimate}} + \underbrace{\sigma_{\sf si}^2
  + \sigma_{\sf down}^2}_{\text{phase noise}}) + \sigma_{\sf noise}^2.
\label{eq:StrengthImperfectBasebandResidue}
\end{equation}

{ \scriptsize
\bibliographystyle{IEEEtran}
\bibliography{FDbib}
}

\begin{table}[ht]
\centering
\scalebox{0.7}{\begin{tabular}{| c | c | } 
\hline  Type of canceller & Expected value of the strength of residual self-interference after active analog cancellation \\ \hline &  \\
  Pre-mixer & $|h_{\sf si}|^2(1 + |\rho|^2 - 2|\rho| R_{x_{\sf si}}(\Delta_{\sf si} -
          \tau) + 2 \sigma_{\sf si}^2(1  - R_{\phi_{\sf si}}(\Delta_{\sf si})) + \sigma_{\sf noise}^2$  \\   &  \\ \hline & \\ 
  Post-mixer & $|h_{\sf si}|^2(1 + |\rho|^2 - 2|\rho| R_{x_{\sf si}}(\Delta_{\sf si} -
          \tau) + 2 \sigma_{\sf si}^2(1  - R_{\phi_{\sf si}}(\Delta_{\sf si} -\tau)) + \sigma_{\sf noise}^2$   \\  &  \\ \hline & \\
  Baseband analog & $|h_{\sf si}|^2 (1 + |\rho|^2 - 2|\rho| R_{x_{\sf si}}(\Delta_{\sf si} - \tau) + (\sigma_{\sf si}^2 + \sigma_{\sf down}^2)) + \sigma_{\sf noise}^2$ \\ & \\ \hline 
\end{tabular}}
\caption{Expected value of the strength of the residual
  self-interference after active analog cancellation with imperfect
  estimate of self-interference channel}
\label{table:ResidualEnergy}
\end{table}

\begin{table}[htb]
\begin{center}
 \scalebox{0.8}{ \begin{tabular}{ |c | c | c | }
    \hline & $\beta_{\phi}^2$ for active analog cancellation only  &  $\gamma_{\phi}^2$ for active analog + digital cancellation  \\ & (imperfect estimate) & (both with imperfect estimates) \\  \hline 
& & \\
 Pre-mixer &
    $2\sigma_{\sf si}^2 (1 - R_{\phi_{\sf si}}(\Delta_{\sf si}))$ & $(1 + |\rho|^2 - 2|\rho| R_{x_{\sf si}}(\tau - \Delta_{\sf si}))(\sigma_{\sf si}^2 + \sigma_{\sf down}^2)$ \\ & & $+ 2\sigma_{\sf si}^2(1 - R_{\phi_{\sf si}}(\Delta_{\sf si}))$\\ 
& & \\ \hline & & \\ Post-mixer & $2\sigma_{\sf si}^2 (1 - R_{\phi_{\sf si}}(\tau - \Delta_{\sf si}))$ & $(1 +|\rho|^2 - 2|\rho| R_{x_{\sf si}}(\tau - \Delta_{\sf si}))(\sigma_{\sf si}^2 + \sigma_{\sf down}^2)$ \\ & & $+2\sigma_{\sf si}^2(1 - R_{\phi_{\sf si}}(\tau - \Delta_{\sf si})) $  \\
 & & \\ 
 \hline & & \\ Baseband canceller & $\sigma_{\sf si}^2 + \sigma_{\sf
      down}^2$ & $\sigma_{\sf si}^2 + \sigma_{\sf down}^2$ \\ & & \\ \hline
 \end{tabular}}
\end{center}
\caption{Parameters defining the singal models in
  \eqref{eq:DiscreteTimeModel1} and \eqref{eq:DiscreteTimeModel2} for
  SISO narrowband, \eqref{eq:DiscreteTimeModel3} for SISO wideband and
  \eqref{eq:MIMOmodel1} for MIMO full-duplex for different types of
  cancellers. We assume that $\sigma_{\sf si} = \sigma_{\sf cancel}$}
\label{table:model}
\end{table}

\end{document}